\documentstyle[emulate_apj,apjfonts,epsf]{article}

\def \farcs{\hbox{$.\!\!^{\prime\prime}$}}
\def \farcm{\hbox{$.\!\!^{\prime}$}}

\lefthead{Hoekstra et al.}
\righthead{Weak lensing analysis of Cl~1358+62}
\begin{document}

\slugcomment{Accepted for publication in the ApJ}
\title{Weak lensing analysis of Cl~1358+62 using {\it Hubble Space Telescope} observations$^1$}

\author{Henk Hoekstra, Marijn Franx, Konrad Kuijken}
\affil{ Kapteyn Astronomical Institute, University of Groningen, \\
        Postbus 800, 9700 AV Groningen, The Netherlands \\
        e-mail: hoekstra, franx, kuijken@astro.rug.nl}
\and 
\author{Gordon Squires}
\affil{ Center for Particle Astrophysics, University of California, \\
        Berkeley, CA 94720 USA \\
        e-mail: squires@magicbean.berkeley.edu}

\begin{abstract}
We report on the detection of weak gravitational lensing of faint,
distant background objects by Cl~1358+62, a rich cluster of galaxies
at a redshift of $z$=0.33. The observations consist of a large,
multi-color mosaic of HST WFPC2 images. 

The number density of approximately 50 background objects
arcmin$^{-2}$ allows us to do a detailed weak lensing analysis
of this cluster. We detect a weak lensing signal out to $\sim$~1.5~Mpc
from the cluster centre. The observed distortion is consistent with a 
singular isothermal sphere model with a velocity dispersion of $780\pm50$ 
km/s. The total projected mass within a radius of 1 Mpc, corresponding to this model
is $(4.4\pm 0.6)\times 10^{14}~{\rm M}_\odot$. The errors given here, represent 
the random error due to the ellipticities of the background galaxies. The uncertainty 
in the redshift distribution introduces an additional, systematic error of 
$\sim 10\%$ in the weak lensing mass. The weak lensing mass is slightly lower than 
dynamical estimates and agrees well with X-ray mass estimates. The mass distribution 
is elongated similar to the light. The axis ratio of $0.30\pm0.15$ and position angle 
of $-21^\circ\pm7^\circ$ were measured directly from the observations and agree very 
well with a previous strong lensing determination. A two-dimensional reconstruction 
of the cluster mass surface density shows that the peak of the mass distribution 
coincides with the peak of the light distribution. We find a value of 
$(90\pm 13)h_{50}{\rm M}_\odot/{\rm L}_{V\odot}$ for the
mass-to-light ratio, consistent with being constant with radius. 

The point spread function of HST is highly anisotropic at the edges of the 
individual chips. This systematically perturbs the shapes of objects and we 
present a method for applying the appropriate correction.

\end{abstract}


\section{Introduction}
\footnotetext[1]{Based on observations with the NASA/ESA {\it Hubble 
Space Telescope} obtained at the Space Telescope Science Institute,
which is operated by the Association of Universities for Research
in Astronomy, Inc., under NASA contract NAS 5-26555}

The technique of weak gravitational lensing has proven to be an important tool 
to study mass distributions in the universe. The projected mass distribution 
of foreground gravitational structures distorts the images of the faint 
background galaxies. As a result gravitational lensing provides a direct 
measurement of the projected mass density (e.g. Kaiser \& Squires 1993).

Until recently, massive structures in the universe were studied
through dynamical analyses of their luminous components. These studies
have shown that large amounts of dark matter exist in the universe.
For clusters of galaxies a popular method uses the motions of the 
galaxies to estimate the mass using the virial theorem. One also can 
estimate the cluster mass profile from X-ray observations when
one assumes hydrostatic equilibrium and spherical symmetry
(e.g. Allen \& Fabian 1994). 

Both methods assume some dynamical state or geometry
in order to obtain the mass or a mass profile. The advantage
of gravitational lensing is the fact that no such assumptions 
are needed. In the regime of weak gravitational lensing one can 
calculate the projected mass surface density up to some additive constant 
from the observed distortion pattern (Kaiser \& Squires 1993; Kaiser et al. 
1994; Schneider \& Seitz 1995; Schneider 1995; Squires \& Kaiser 1996).

Since the first succesful measurements of the weak gravitational distortions
(Tyson, Valdes \& Wenk 1990), many massive clusters of galaxies have been 
studied (e.g. Bonnet, Mellier, \& Fort 1994; Fahlman et al. 1994; Squires et
al. 1996b; Luppino \& Kaiser 1997). In principle one can measure the 
gravitational distortion out to large radii from the cluster centre, beyond 
the radii where X-ray observations or cluster kinematics can be used to
determine the mass distribution. 

So far, most weak lensing studies of clusters of galaxies have been
undertaken using data from ground based telescopes. These are affected
by atmospheric seeing, which causes the images of the faint background
galaxies to be enlarged and more circular. 

In this paper we present the first weak lensing analysis of a cluster of
galaxies using HST observations with a large field of view. Up to now, other 
weak lensing studies of clusters of galaxies with HST have been limited to 
cluster cores (C.~Seitz et al. 1996; Smail et al. 1997). These observations 
consisted of single pointings, thus suffering from the limited field of view 
of HST.

Our data consist of a mosaic of 12 pointings, thus yielding a total field of 
view of approximately 8 by 8 arcmin. This combination of space based
observations and a large field of view provides an interesting
opportunity to study the cluster Cl~1358+62 in great detail.
The HST observations have also been used to study the evolution of cluster
galaxies as a function of redshift (Kelson et al. 1997;
van Dokkum et al. 1998).

Furthermore the HST observations revealed a giant red arc approximately
$21''$ from the central galaxy. Franx et al. (1997) showed
that the arc is a gravitationally lensed image of a galaxy at
a redshift of 4.92. A strong lensing model based on this arc and 
its counterarc yields a velocity dispersion of 970 km/s, which
is in fair agreement with the cluster kinematics (Franx et al. 1997).

An advantage of HST observations is the high number density of galaxies one
can reach in relatively short exposures. Previous HST studies have achieved
$\simeq 100$~galaxies arcmin$^{-2}$ routinely (e.g. C.~Seitz et al, 1996;
Smail et al. 1997). With only 3600 seconds exposures per pointing, we
obtain a number density of $\sim 50$ useful background galaxies arcmin$^{-2}$ 
in both F606W and F814W.

Another important advantage of HST over ground based observations is the size
of the point spread function (PSF). Most of the faint objects are small. To 
recover the lensing signal one  needs to correct for the effect of seeing 
(Bonnet \& Mellier 1995; Kaiser, Squires \& Broadhurst 1995 (KSB95 hereafter); 
Luppino \& Kaiser 1997 (LK97 hereafter); Fischer \& Tyson 1997). For objects 
with sizes comparable to the PSF, these corrections become very large, 
amplifying the uncertainty in the ellipticity due to photon noise. As a 
result the scatter in the derived ellipticities of the galaxies is larger 
than the expected scatter due to their intrinsic shapes. Consequently, for 
a given number density of background objects, the accuracy of weak lensing 
studies based on HST observations will be higher than the results from ground 
based data.

We will investigate the dark matter and galaxy distribution
in the cluster of galaxies Cl~1358+62. The cluster is at a
redshift of 0.33 with a measured velocity dispersions of $1027^{+51}_{-45}$ 
km/s (Fisher et al. 1998) and $910\pm54$ km/s (Carlberg et al. 1997).  It is 
also a rich cluster at its redshift, with Abell richness class 4 (Luppino et 
al. 1991). The cluster has a strong binary nature (Carlberg et al. 1997) with 
a massive substructure moving at $\sim 1000$~km/s with respect to the cluster 
centre. This substructure is outside the region observed in this paper.
Cl~1358 has a measured X-ray luminosity of
$L_x~(0.2-4.5~{\rm keV})~=~7 \times 10^{44} h_{50}^{-2}~{\rm erg/s}$
(Bautz et al. 1997) which makes it a luminous cluster in 
X-ray\footnote{Throughout this paper we will use $h_{50}=H_0/
(50~{\rm km/s/Mpc})$, $q_0$=0.5 and $\Lambda=0$. This gives a scale 
of $1''=5.8~h_{50}^{-1}$ kpc at the distance of Cl~1358.}. 

The observations and data reduction are briefly outlined in section~2.
In section~3 we discuss the method we used, including the
corrections for both the anisotropy and the size of the point spread
function, and the camera distortion. The object selection for the weak 
lensing analysis is described in section~4. The light distribution
is discussed in section~5. The redshift distribution we used to calculate
the critical surface density are presented in section~6. The results from the 
weak lensing analysis are presented in section~7. In the appendices we 
address the issue of PSF and camera distortion corrections.
   
\section{Data}

The cluster Cl~1358+62 was observed with the WFPC2 camera on board the 
Hubble Space Telescope in February 1996. The observations consist
of 12 pointings of the telescope. Each pointing consists of three
exposures of 1200s in both the F606W and F814W filter.
Combining the observations yields a mosaic of approximately 8 by 8
arcminutes, which makes it among the largest fields observed with HST.
The total area covered by the observations is approximately 53 arcmin$^2$.

The data reduction is described in van Dokkum et al. (1998). 
For the weak lensing analysis we omit the data of the PC chips because
of the brighter isophotal limit. 

\section{Method}

Our analysis technique is based on that developed by KSB95, with a
number of important modifications. We summarize the method here in
order to highlight the differences, which center on the correction of
certain arithmetic errors in KSB95's formulae, and on a careful study
of the effect of the weight function used in calculating image
moments. The details of the complete method are given in the
appendices.

The first step in the analysis is to detect the faint galaxy images,
which we do with the KSB95 algorithm and software. The shapes of these
objects are then quantified by calculating the central second moments $I_{ij}$
of the image fluxes and forming the two-component polarization (Blandford et 
al. 1991)

\begin{equation}
e_1 = \frac{I_{11}-I_{22}}{I_{11}+I_{22}}~{\rm and}
~e_2 = \frac{2I_{12}}{I_{11}+I_{22}}
\end{equation}
Because of photon noise, unweighted second moments cannot be used.
Instead, we use a circular gaussian weight function with
dispersion $r_g$ equal to the radius of maximum significance given by
the KSB95 detection algorithm.

Before we can search for lensing-induced systematics in the shapes of
the objects we have detected, other sources of distortion have to be
corrected for. The relevant one for HST observations is smearing by
the PSF, which may be viewed as two separate effects: the effect of
the anisotropy of the WFPC2 PSF (see below), which will cause a
systematic polarization of galaxy images, and the circularization of
the galaxy images through convolution with the isotropic part.

We investigated the point spread function of WFPC2 using stars
from our observations. Using these stars we found indications of
a significant anisotropy in the PSF. The results are shown in 
Figure~\ref{starmos}. The orientation of the sticks indicates the direction of 
the major axis of the stars, and the size is proportional to the
size of the anisotropy.

Figure~\ref{starmos} clearly shows that the anisotropy changes as
a function of position and becomes larger towards the edges of the
chips. Unfortunately the limited number of stars in our observations 
does not provide sufficient coverage of the chips. One also
finds from this figure that the patterns in the two filters are 
fairly similar.

To investigate the PSF anisotropy as a function of position 
in more detail we retrieved  observations of the globular cluster M~4 
(Richer et al. 1997) from the HST archive. The data consist of a pointing 
at the core and two pointings at one effective radius. The polarization 
pattern we measured from isolated stars in this globular cluster is shown 
in Figure~\ref{starglob}. 

Figure~\ref{starglob} shows that especially at the edges of the chips the
anisotropies are large. According to Holtzman et al. (1995a) the PSF changes 
with field position because of a variable pupil function and small 
aberrations. The anisotropy introduced by the camera distortion is discussed 
below. Also variations in time occur due to jitter and focus changes. 
Nevertheless, we use the M~4 data as a starting point for the PSF modeling, 
and correct for residual differences as described below.

\vbox{
\begin{center}
\leavevmode
\hbox{%
\epsfxsize=8cm
\epsffile{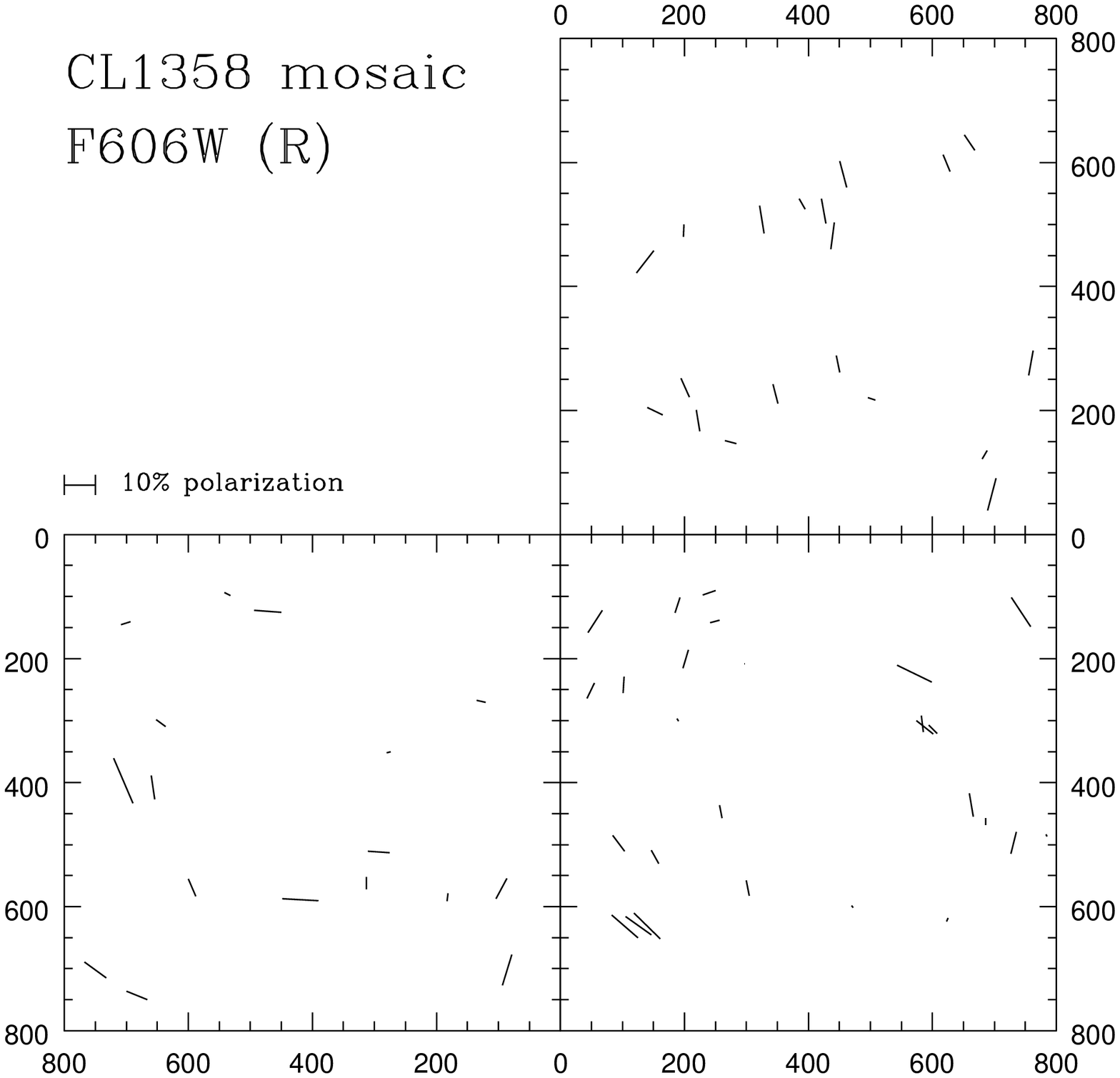}}
\hbox{%
\epsfxsize=8cm
\epsffile{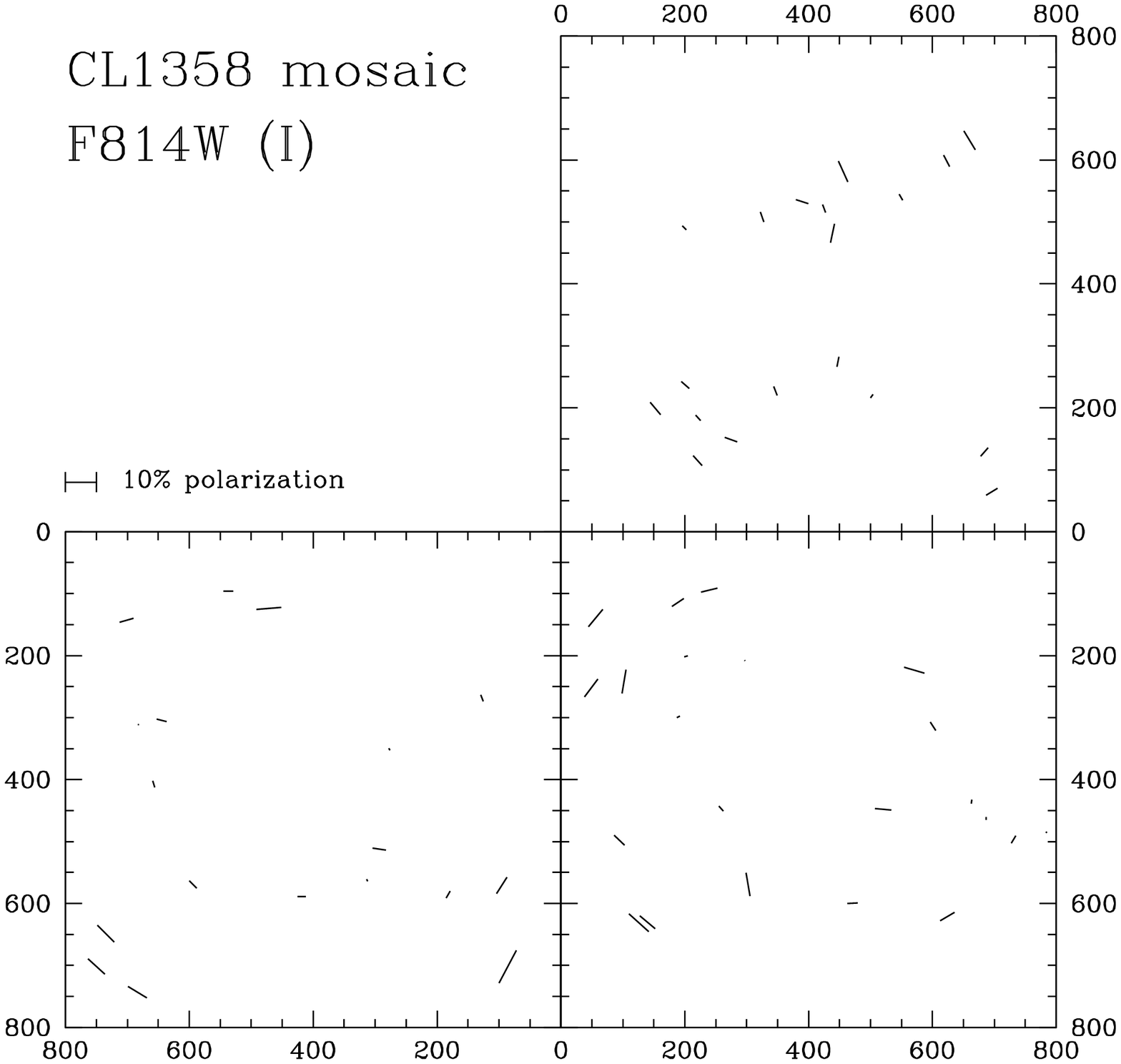}}
\begin{small}
\figcaption{The polarization field of stars taken from observations of the 
cluster of galaxies Cl~1358+62. The orientation of the sticks shows the
direction of the major axis of the PSF whereas the length is
proportional to the size of the anisotropy. The polarization are
calculated using a Gaussian weight function with a dispersion of 1 pixel,
which is the value found by the peak finder. The upper figure shows the 
results from the F606W data. The lower figure shows the results from the 
F814W data. The lower left panel is chip~2, lower right is chip~3 and the 
upper right one denotes chip~4. We have omitted chip~1, which is the 
planetary camera.\label{starmos}}
\end{small}
\end{center}}

In the original KSB95 approach only one size weight function was used for 
the anisotropy model. This is generally sufficient
for ground based data, for which the PSF can be well represented by a 
Gaussian. We have found that the very non-Gaussian WFPC2 PSF requires that 
we use correction parameters which depend on the size of the object we want 
to correct. 

\vbox{
\begin{center}
\leavevmode
\hbox{%
\epsfxsize=8cm
\epsffile{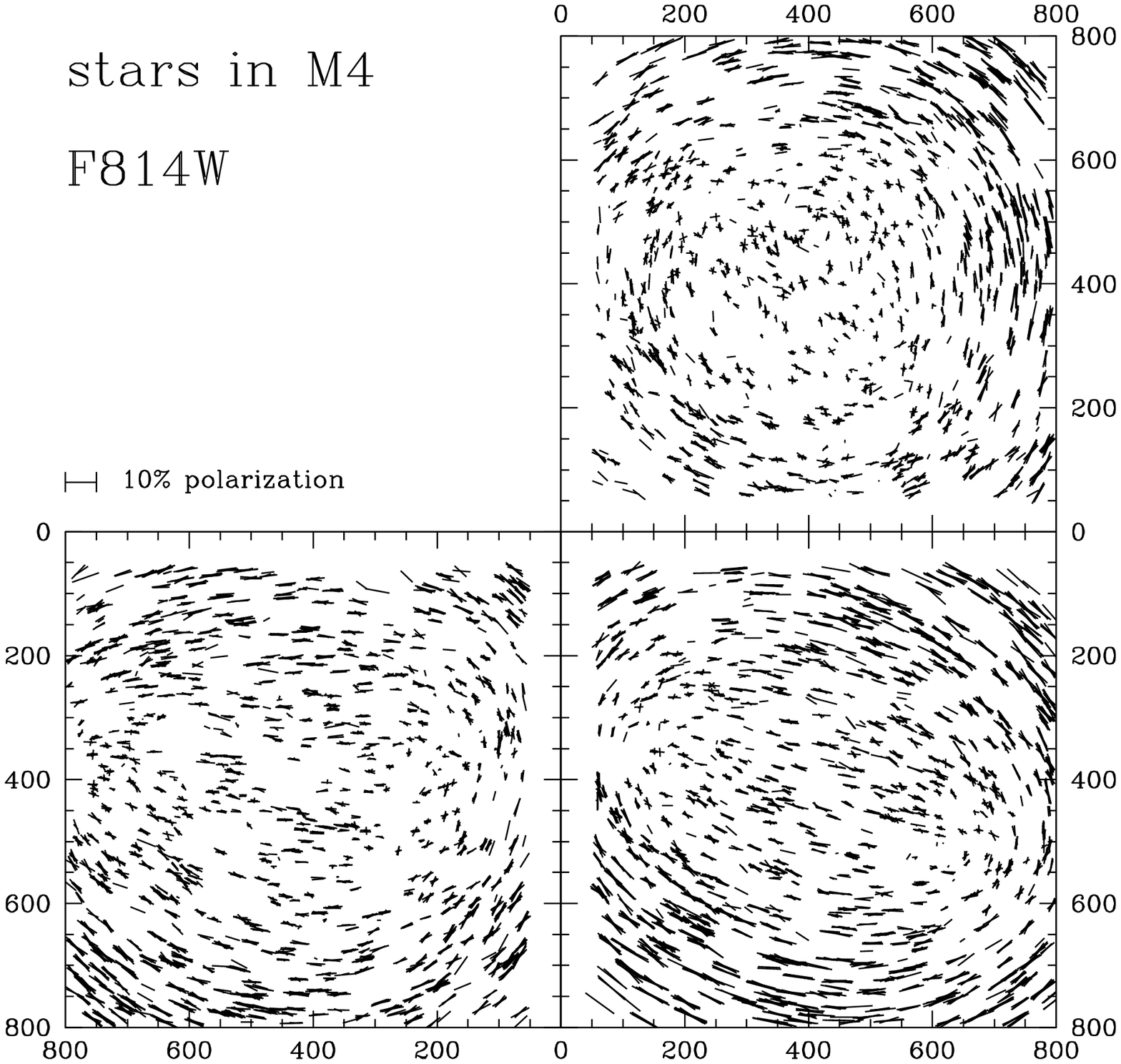}}
\hbox{%
\epsfxsize=8cm
\epsffile{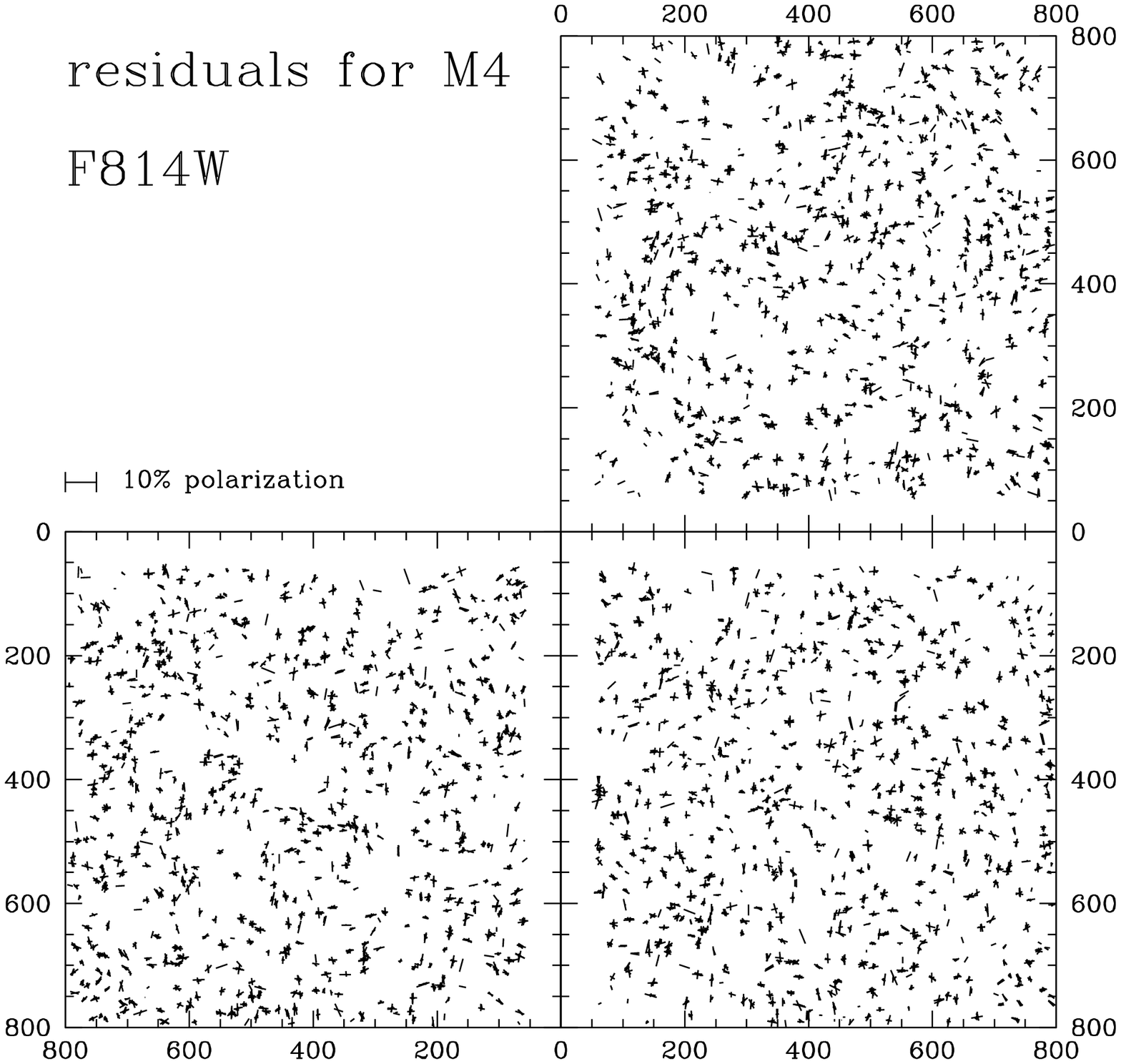}}
\begin{small}
\figcaption{The upper panel shows the  polarization field of stars taken 
from observations of the globular cluster M~4. The orientation of the 
sticks shows the direction of the major axis of the PSF whereas 
the length is proportional to the size of the anisotropy. These 
observations were  taken in the F814W filter. The polarizations are
calculated using a Gaussian weight function with a dispersion 1 pixel,
which is the value given by the peak finder. The lower left panel is chip~2, 
lower right is chip~3 and the upper right one denotes chip~4. We have omitted 
chip~1, which is the planetary camera. The lower panel shows the residuals
after subtracting a third order polynomial fit for each polarization component
(10 parameters for each component).
\label{starglob}}
\end{small}
\end{center}}

We fit a third order polynomial to the polarization field  of the 
globular cluster for a given width $r_g$ of the weight function. This yields 
a series of maps of the PSF anisotropy as a function of $r_g$, which we will 
use to correct the measured galaxy image shapes. 

The residuals after subtracting the model from the data (using a weight 
function with a dispersion of 1 pixel) are shown in the lower panel of 
Figure~\ref{starglob}. In this figure no systematic patterns are seen and 
a higher order fit provided no significant improvement. 

Because the PSF can change with time it is important to know how
stable the pattern shown in Figure~\ref{starglob} is. To investigate this, 
we retrieved archival HST data for another globular cluster (NGC~6752) in the 
F814W band and compared the results found for that cluster to the results 
found for M~4. We also compared the results from M~4 to the measurements of 
the stars in the mosaic. In both cases we found a fair agreement although
some systematic deviations from the model are seen. These deviations
are on the few percent level. 

To correct for these systematic differences between the observations 
of M~4 and our mosaic, one can fit a low order polynomial to the residuals
between the model and the mosaic. Due to low number of stars per individual
pointing, we stack the results of all 12 pointings for each individual chip.
We found that for our observations a zeroth order polynomial of
$\le 1\%$ is sufficient. We then repeat the procedure for different 
choices of the weight function radius $r_g$, finally ending up with maps at a 
number of values of $r_g$ of the PSF polarizations and polarizabilities for
the three WF chips.

\vbox{
\begin{center}
\leavevmode
\hbox{%
\epsfxsize=8cm
\epsffile{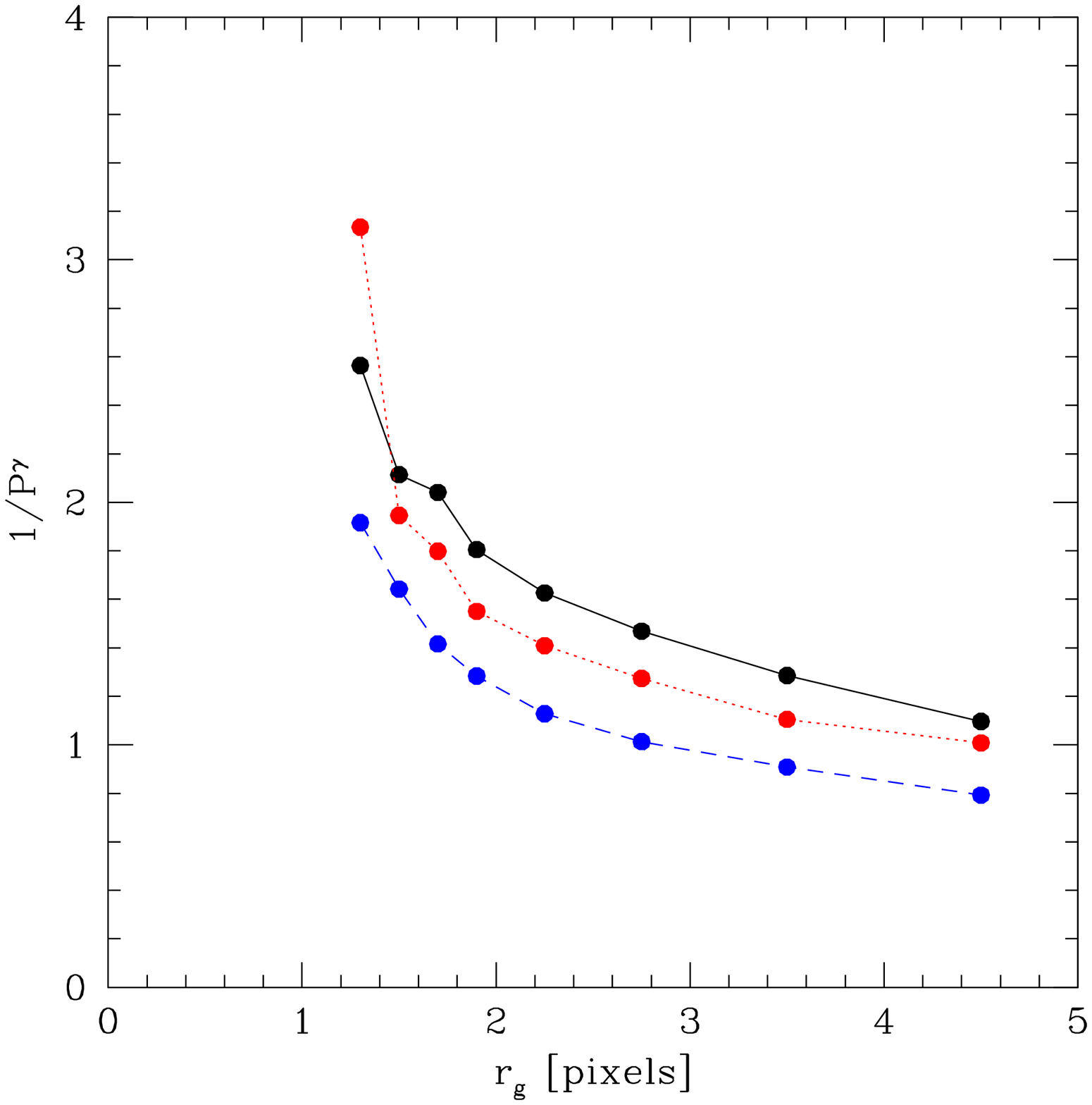}}
\begin{small}
\figcaption{Correction factor when one goes from polarization to
distortion as a function of the radius of maximum significance for
the F814W data. The solid line shows the results for objects
with $18<$F814W$<22$, the dotted line for objects with $22<$F814W$<24$ and
the dashed line for objects with $24<$F814W$<26.5$.
This correction factor is the inverse of the 'pre-seeing'
shear polarizability.\label{corfac}}
\end{small}
\end{center}}

The effect of an anisotropic PSF on the polarization $e_\beta$ of a
galaxy image is quantified by the `smear polarizability' $P^{\rm sm}$,
which measures the response of the image polarization $e_\alpha$ to a
convolution with a small anisotropic kernel. $P^{\rm sm}$ can be
estimated for each observed image, and also depends on the weight
function used in calculating $e_\alpha$. In the appendix, we correct a
small analytical error in the expression for $P^{\rm sm}$ given by
KSB95. Correcting the observed galaxy polarizations using
\begin{equation} 
e_\alpha\to e_\alpha - \sum_\beta {P_{\alpha\beta}^{\rm sm}\over
P_{\beta\beta}^{\rm sm}*}e_\beta*,
\end{equation}
where starred quantities refer to parameters measured for stellar
images, then undoes the effect of PSF anisotropy. In our analysis we
take care to use the same radius for the Gaussian weight function for
all parameters in this equation, though it may differ from galaxy to
galaxy---unlike KSB95, who use a smaller $r_g$ for the stars than for
the galaxies. This turns out to be significant for the WFPC2 PSF, as we
justify and discuss in the appendices. In practice, we interpolate the
map of stellar polarizations and polarizabilities to $r_g$ equal to
each galaxy's radius of maximum significance.

The next step is to deduce the distortion from the anisotropy-corrected
galaxy polarizations. Both the (now effectively isotropic) PSF, and
the circular weight function, tend to make objects rounder. These
effects may be corrected for using the `pre-seeing shear
polarizability' $P^\gamma$ of LK97. This quantity can again be
estimated from the observed galaxy and star images (in the appendix 
we correct a small analytical error in the expression for the shear 
polarizability in KSB95), and again we have found that it is
important to use the same value of $r_g$ for star and galaxy
polarizations and polarizabilities when constructing $P^\gamma$, as
discussed further in the appendices.

\vbox{
\begin{center}
\leavevmode
\hbox{%
\epsfxsize=8cm
\epsffile{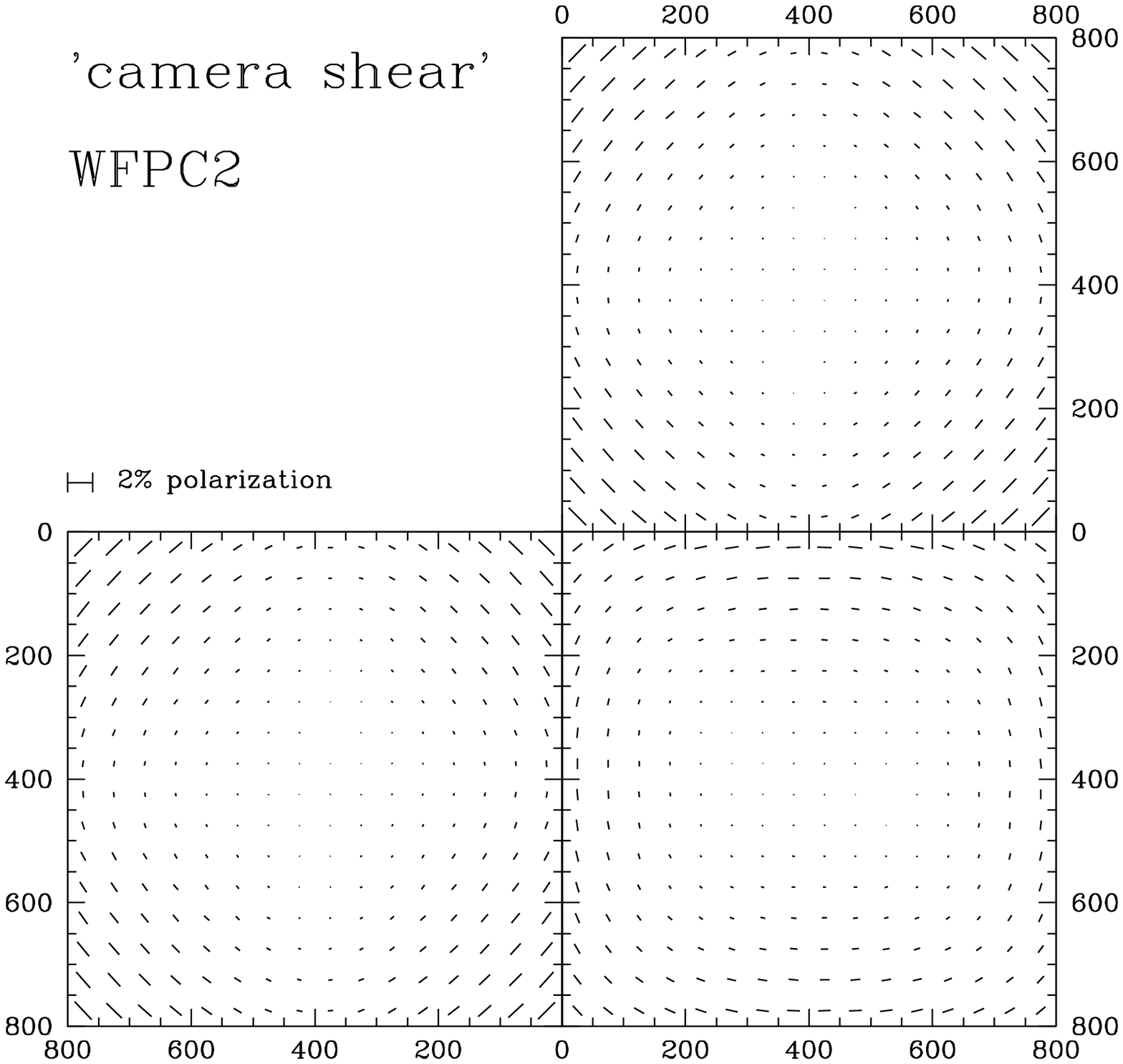}}
\begin{small}
\figcaption{The shear introduced by the WFPC camera distortion. The
shear was calculated using the coefficients for the distortion 
given in Holtzman et al. (1995a). The orientation of the sticks 
shows the direction of the shear and the length is proportional 
to the size. The lower left panel is chip~2, lower right is chip~3 
and the upper right one denotes chip~4. We have omitted chip~1, 
which is the planetary camera. 
\label{camshear}}
\end{small}
\end{center}}

Because of the noisiness of invidual estimates for $P^\gamma$, we determine 
the mean $P^\gamma$ for a series of bins of magnitude and $r_g$. These mean 
values are used to estimate the distortion. In Figure~\ref{corfac} we show the 
correction factor when going from a polarization to a distortion as a function 
of the radius of maximum significance and brightness of the object. The 
correction converges to a value of $\sim$ 1 for large objects. The smallest 
objects have a correction factor that is twice as large. For ground based 
data much higher values would be found. Figure~\ref{corfac} also shows 
that the correction factor increases with increasing brightness, indicating
that these objects have on average different profiles than fainter objects.
Using the original equations from KSB95, would underestimate the shear
polarizability by approximately 13\%, therefore overestimating the distortion 
by a similar amount.

\vbox{
\begin{center}
\leavevmode
\hbox{%
\epsfxsize=7cm
\epsffile{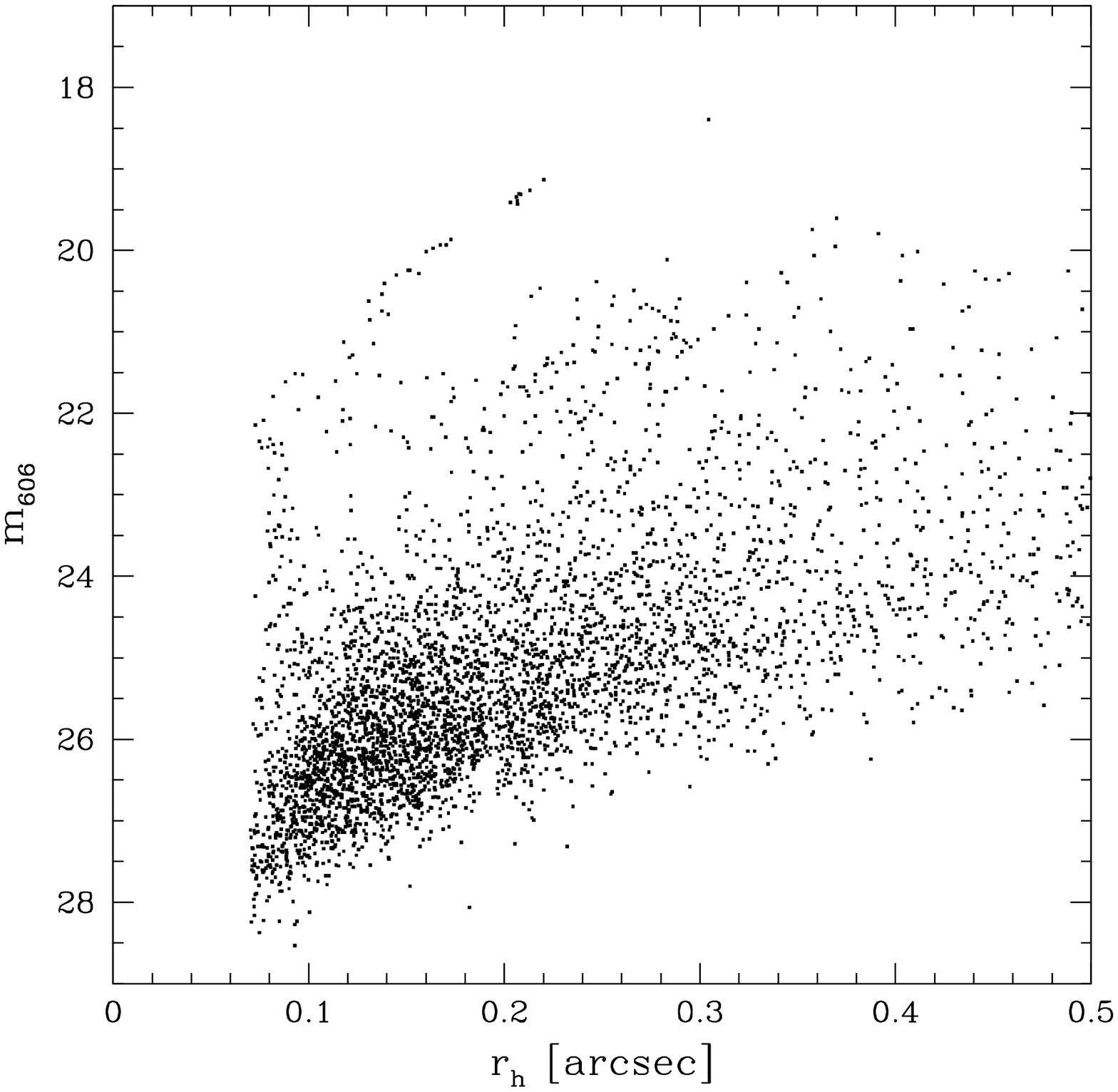}}
\hbox{%
\epsfxsize=7cm
\epsffile{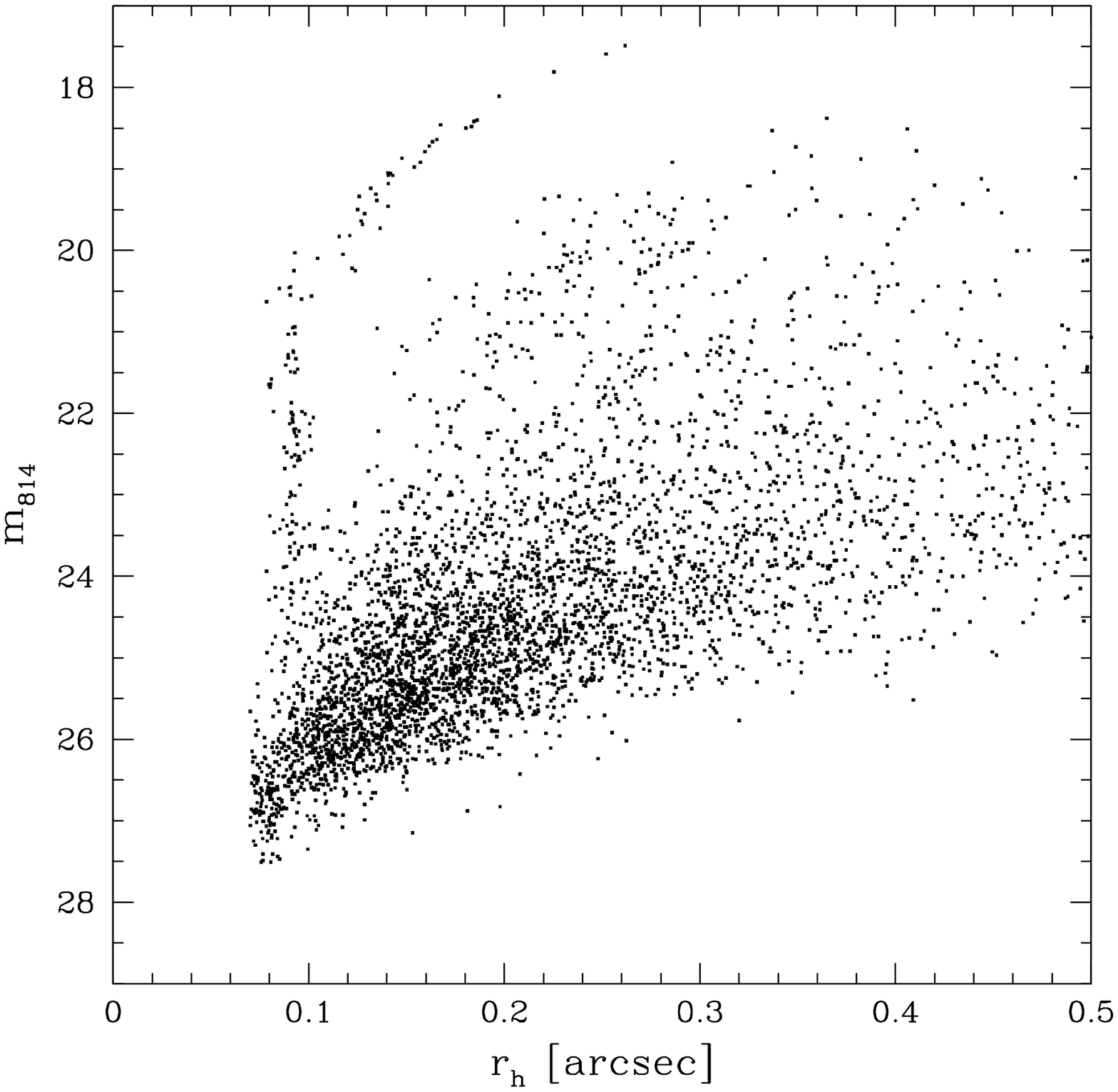}}
\begin{small}
\figcaption{Plots of the magnitude of selected objects versus the 
calculated half light radius. The upper plot shows the results
for the F606W data. The lower plot shows the results of the 
F814W data. In both figures the vertical stellar locus is
clearly seen. We identify galaxies as objects that have a
half light radius that is 1.2 times that of the stars.\label{coin}}
\end{small}
\end{center}}

\noindent We finally estimate the distortion $g$ at a certain position using
\begin{equation} 
g_\alpha=\frac{\langle e_\alpha \rangle}{\langle 
P^\gamma_{\alpha\alpha} \rangle}.
\end{equation}

So far we have neglected the effect of the camera distortion. A distortion
introduced by the camera mimics a shear. We show in appendix~A2 that the
correction for this effect is straightforward. To obtain the true shear
one just has to subtract the camera shear from the observed shear in 
equation~3.

The distortion of WFPC2 is described in Holtzman et al. (1995a). We used the 
coefficients from their paper to calculate the shear induced by the camera
distortion. The resulting shear field is presented in Figure~\ref{camshear}.
The effect is small; the largest values are 2\% in the corners of the chips.
This shear field is used to correct the observed distortion.

The HST point spread function is badly sampled. As a result it introduces 
some extra scatter in the measurement of $e_1$. To investigate how the 
correction scheme is affected by the sampling problem we used simulations, 
which are described in the appendices. The simulations show that the 
corrections can be applied safely to objects that have a radius of maximum 
significance of 1.2 pixel and larger.

\section{Object selection}

The cluster was observed in two filters: F606W and F814W. 
For each pointing and filter, we detected objects with a significance
of 4$\sigma$ over the local sky, using the peak finder described in
KSB95. Only objects that were detected in both the F606W and F814W exposure
were used for further analysis. Though it provides an efficient way to
remove spurious detection, a disadvantage of this approach is that it also
removes faint red objects (only detected in F814W) and faint blue objects 
(only detected in F606W).  Objects that were detected both in F606W and 
F814W but clearly did not match in size were also removed. 

\begin{figure*}
\begin{center}
\leavevmode
\hbox{%
\epsfxsize=\hsize
\epsffile[18 144 592 440]{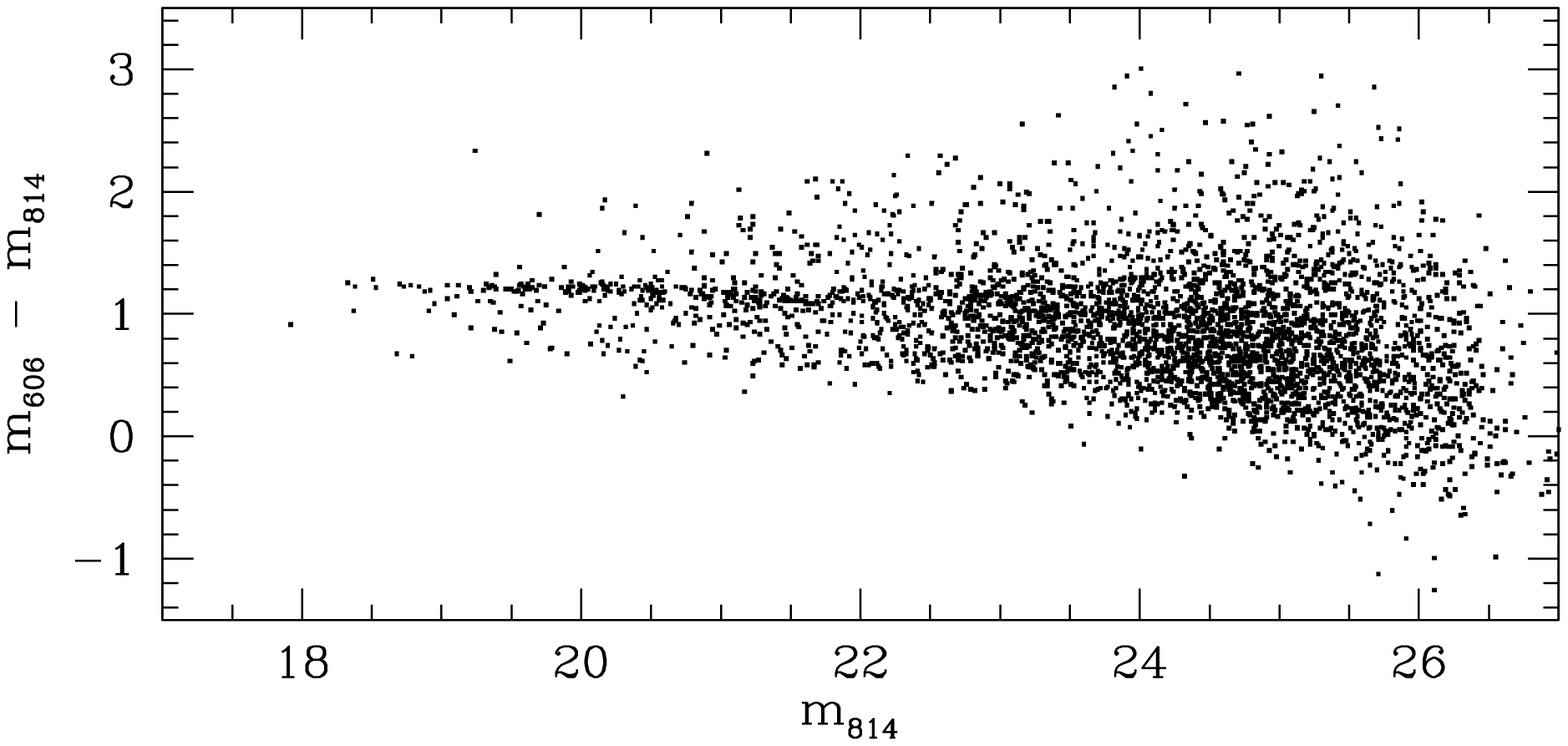}}
\begin{small}
\figcaption{Color-magnitude diagram for objects detected both in F606W
and F814W. In this figure all 4175 objects we identified as being 
galaxies are plotted. Also notice the very sharp cluster color-magnitude 
relation.\label{colmag}}
\end{small}
\end{center}
\end{figure*}

We convert the measured counts to F606W and F814W magnitudes, zero-pointed
to Vega, using the conversions given in Holtzman et al. (1995b).
We add 0.05 magnitude to the zeropoints to account for the CTE effect.
Figure~\ref{coin} shows a plot of the magnitude of objects versus
the calculated half light radius. Stars are located in the vertical
locus in this plot. Bright stars saturate and their measured sizes 
increase. This figure shows that one can separate moderately
bright stars from galaxies and saturated stars. The stars we
identified this way were used to examine the PSF anisotropy
(cf. Figure~\ref{starmos}).

\begin{figure*}
\begin{center}
\leavevmode
\vspace{20cm}
\begin{small}
\figcaption{(a) Luminosity weighted distribution of cluster galaxies
brighter than $V_z$ of 22.5 magnitude; (b) Number density distribution 
of this sample of cluster galaxies; (c) Number density distribution 
of all detected galaxies brighter than $F814W$ 22.5; (d) Number
density distribution of all detected galaxies fainter than $F814W$ 22.5.
The distributions have been smoothed with a Gaussian filter with scale
$0\farcm4$; the shaded circle indicates the FWHM of the smoothing function. 
The contours are 10\%, 20\%, etc. of the peak  value. The total size of the 
image is 500 by 530 arcseconds, within which the mosaic covers the indicated 
region. The orientation is such that north is up and east is to the left.
\label{lumdis}}
\end{small}
\end{center}
\end{figure*}

We select objects with half light radii 1.2 times the half light radii from 
the PSF as galaxies. From this sample we removed objects with $r_g$ less
than 1.2 pixels, as we can correct only galaxies which have a radius of 
maximum significance of 1.2 pixels or more reliably. This is demonstrated
in the appendices.

This yields a catalog of 4175 objects corresponding to a number density of 
79 galaxies arcmin$^{-2}$. Due to our selection criteria, the total
number of detected galaxies is slightly higher. 

Combining the F606W and F814W observations, we determined
the colors of galaxies in our catalog. The resulting color-magnitude 
diagram is shown in Figure~\ref{colmag}. This figure clearly shows the sharp
color-magnitude relation of Cl~1358, which is discussed in van Dokkum et al. 
(1998).

\vbox{
\begin{center}
\begin{tabular}{lcccc}
\hline
\hline
name            & $m_{814}$   & $m_{606}-m_{814}$  & \# galaxies & $\bar n$\\
\hline
bright          & $22-25  $ & $\notin[0.9,1.5]$  & 1272 & 24  \\
faint           & $25-26.5$ & $--$               & 1392 & 26  \\ 
blue            & $22-26.5$ & $<0.9$             & 1835 & 35  \\
red             & $22-26.5$ & $>1.5$             & 393  & 7   \\
\hline
\hline
\end{tabular}
\end{center}
\begin{small}
{\sc Table~1.} Properties of the various subsamples taken from
the final catalog of galaxies.
\vspace{0.4cm}
\end{small}}

The sharp cluster color-magnitude relation allows us to remove many bright 
cluster members from our catalog. At the faint end the cluster sequence 
becomes broader and blends with the population of background galaxies. 
The color information allows us to create color selected subsamples of 
background galaxies. The subsamples we use in the weak lensing analysis 
are listed in Table~1.

\section{Light distribution}

We calculated the luminosity in the redshifted $V$ band as described
in van Dokkum et al. (1998). The direct transformation from the HST
filters to the redshifted $V$ band is given by
$$ V_z = m_{814} + 0.2 (m_{606}-m_{814}) + 0.65,$$
\noindent where $V_z$ denotes the redshifted $V$ band magnitude.
The luminosity in the redshifted $V$ band is 
$$ L_V=10^{0.4(M_{V\odot}-V_z+DM+A_{F814W})}~L_{V\odot},$$
where $M_{V\odot}=4.83$ is the solar absolute $V$ magnitude, $DM$ is the
distance modulus, and $A_{F814W}$ is the extinction correction in the F814W 
band. The redshift of 0.33 gives a distance modulus for Cl~1358 of 
$41.63-5\log h_{50}$ (or $41.52-5\log h_{50}$ using $q_0=0.1$). 
Taking galactic extinctions from Burstein \& Heiles (1982) and using the 
results from Cardelli, Clayton, \& Mathis (1989) we find a value of 0.02 
for $A_{F814W}$.

To measure the total luminosity of the cluster galaxies, we measured
aperture magnitudes using apertures of 3 arcsecond diameter.
Confirmed cluster galaxies were fitted by $\exp(r^{1/n})$ profiles
by Pieter van Dokkum. Using the total magnitudes inferred from these
fits, we estimate the aperture correction as a function of aperture
magnitude.

Many galaxies in the Cl~1358 field were observed spectroscopically 
(Fisher et al. 1998). Down to $V_z=20.5$ the catalog of confirmed 
cluster members is complete. To estimate the light of the cluster
galaxies we use the confirmed cluster galaxies if they are
brighter than $V_z=20.5$ ($F814W\approx 19.6$). For galaxies with 
$20.5<V_z<22.5$ we select galaxies that lie on or less than 0.2 
magnitude below the color magnitude relation, yielding a catalog 
that contains 341 galaxies. At even fainter magnitudes the color
magnitude relation blends with the population of background galaxies.

Using the sample of bright cluster galaxies, we estimate a 
luminosity function. We found that a Schechter function with $\alpha=-1.1$ 
and ${\rm L}_*=4.5\times 10^{10}{\rm L}_{V\odot}$ fits the observed counts 
well. From this luminosity function we estimate that galaxies fainter than 
$V_z=22.5$ contribute $11\%$ to the total light of the cluster. 

We calculate the cumulative light profile from the sample of bright
cluster galaxies. The profile is multiplied by a factor 1.11 to
account for the light from faint cluster galaxies. The cumulative light 
profile is shown in Figure~\ref{cumlum}. The solid line in this
figure corresponds to an isothermal profile, indicating that
the radial light profile is close to isothermal. The total 
luminosity within an aperture of 1~Mpc is 
$(5.19\pm0.26)\times 10^{12}{\rm L}_{V\odot}$.

To obtain a qualitative description of the cluster light distribution, 
we calculated both the galaxy number density and the galaxy light.
Grey scale plots of the luminosity and number weighed light distributions 
of the sample of bright cluster galaxies are presented in Figure~\ref{lumdis}. 
Both distributions have been smoothed with a Gaussian filter with scale 
$0\farcm4$. The galaxy number density and the luminosity weighted 
distributions of the cluster sample look quite similar. The dominant 
structure in both distributions coincides with the region
around the central galaxy in the cluster. Both figures show that the
cluster light is elongated north-south. 
Several other concentrations are also visible. Of these the one
to the north is the most significant. The extension to the south and the
concentration to the north were already reported in Luppino et al. (1991).

To investigate the clustering of background galaxies, we also show 
the number density distribution of all detected galaxies. We split
the sample into two bins. Figure~\ref{lumdis}c shows the number density
distribution of galaxies with F814W brighter than 22.5, whereas 
Fig.~\ref{lumdis}d shows the result for galaxies fainter than F814W
magnitude 22.5.

The counts in bright sample are dominated by the cluster. For the faint
galaxies we see two peaks near the edges of the observed region. As these
enhancements in the number density are not seen in the bright sample,
they correspond to galaxy concentrations at redshifts higher than the
redshift of Cl~1358.

\vbox{
\begin{center}
\leavevmode
\hbox{%
\epsfxsize=8cm
\epsffile{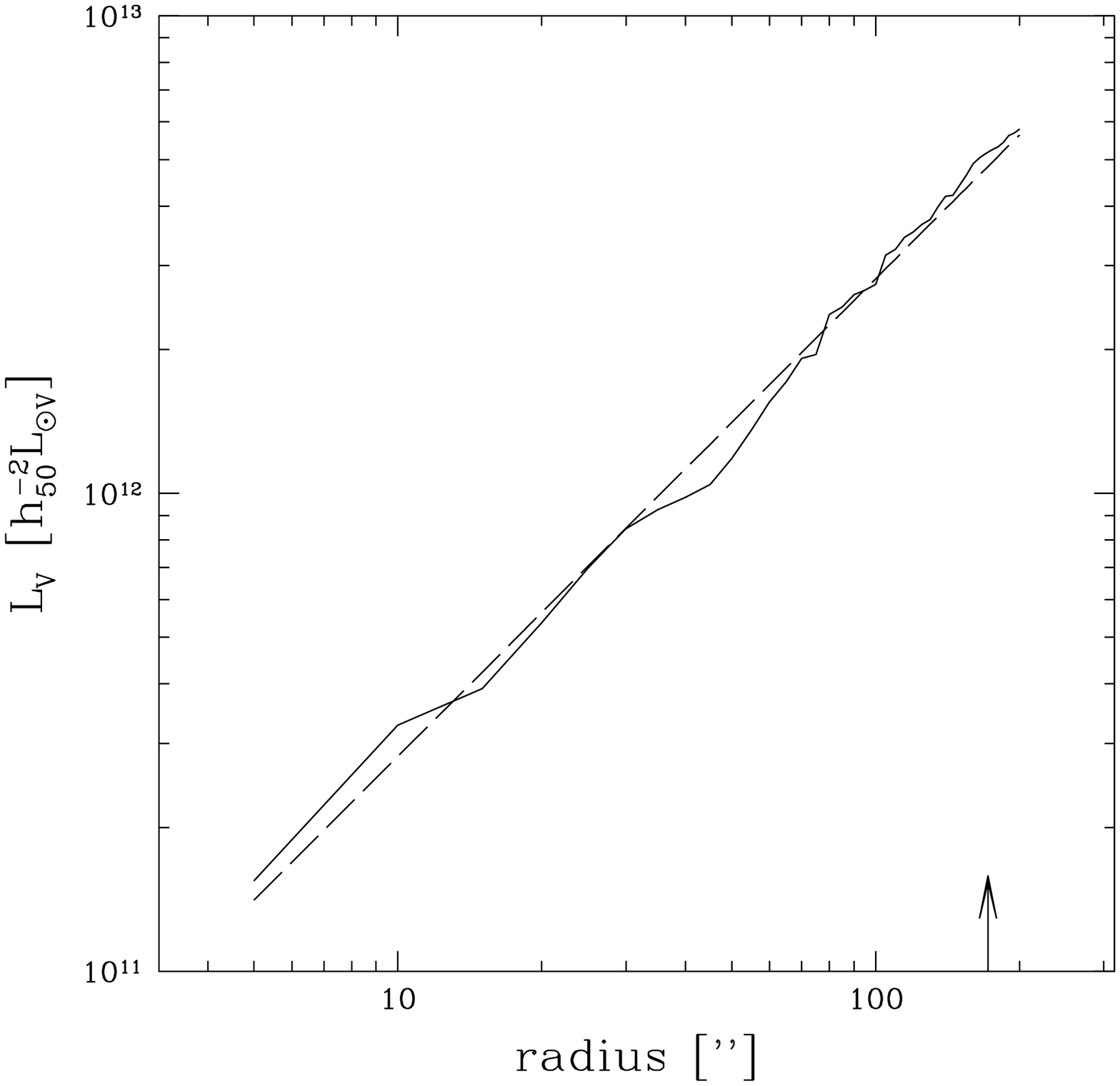}}
\begin{small}
\figcaption{The cumulated, passband corrected, $L_V$ luminosity as 
a function of radius from the central elliptical galaxy. The dashed line 
corresponds to a isothermal profile. The arrow indicates a radius of 
1~$h_{50}^{-1}$ Mpc.\label{cumlum}}
\end{small}
\end{center}}

\section{Redshift distributions}

From the measured distortion one can calculate the dimensionless
surface density $\kappa$ (e.g. Schneider, Ehlers \& Falco 1992)
up to some additive constant.
To convert this dimensionless surface density into a physical 
surface density we have to estimate a mean critical surface density
$$\Sigma_{\rm crit}=c^2 (4 \pi G D_l \beta)^{-1},$$
where $D_l$ is the angular diameter distance to the lens and
$$\beta=\left\langle {\rm max}\left(0,1-\frac{1-(1+z_l)
^{-{\frac{1}{2}}}}{1-(1+z_s)^{-{\frac{1}{2}}}}\right)\right\rangle$$ 
in an Einstein-De Sitter universe ($\Omega_0=1$) when $z_l$ and $z_s$ are the 
redshifts of resp. lens and source. $\beta$ depends on the redshift 
distribution of the background galaxies.
The redshift of Cl~1358 gives an angular diameter distance of 
$1.20~h_{50}^{-1}~{\rm Gpc}$, yielding a value of 
$\Sigma_{\rm crit}=1380\beta^{-1}~h_{50}~{\rm M}_\odot {\rm pc}^{-2}$
(using $q_0=0.1$, the angular diameter distance to Cl~1358 is 
$1.14~h_{50}^{-1}~{\rm Gpc}$).

Unfortunately the redshift distribution of the faint galaxies
that are used in this weak lensing analysis is uncertain. Down
to a magnitude of $I\sim 23$ spectroscopic redshift surveys
give a fairly good picture of the redshift distribution. At
fainter magnitudes redshifts can be estimated by broad-band
photometric redshift techniques (e.g. Sawicki, Lin, \& Yee 1997;
Lanzetta, Yahil, \& Fern\'andez-Soto 1996; Lanzetta, Fern\'andez-Soto, 
\& Yahil 1997), although the reliability of these techniques is still 
uncertain.

Another approach is to model galaxy evolution and estimate
redshift distributions from the model predictions. The
results are uncertain and the distributions at faint magnitudes can differ much
between various models.

We used $n(m,z)$ distributions provided by Caryl Gronwall 
(Gronwall \& Koo 1995), Rychard Bouwens and Joseph Silk 
(Bouwens \& Silk, 1996; Bouwens et al. 1997), and Harry Ferguson 
(Babul \& Ferguson 1996) to estimate $\beta$ for our galaxy samples. 
The Gronwall-Koo model and the Babul-Ferguson model we have were calculated 
for a flat universe. The low $\Omega$ Pozzetti model 
(Pozzetti, Bruzual, \& Zamorani 1996) was calculated by Rychard Bouwens.

If available, we used low $\Omega$ models, as our results (cf. section~7.3) are
consistent with a low density universe. Due to the relatively low redshift
of Cl~1358 and the relatively high redshifts of the background galaxies, 
the value for $\beta$ does not change whether one uses $q_0=0.5$ or $q_0=0.1$.

Another problem arises for the faint sample due to the size
cut that was needed to get reliable corrections for the PSF
effects. Such a size cut will change the redshift distribution
of the objects in our faint sample and therefore increases the
uncertainty in $\beta$. The effect of the size cut is negligible for the 
bright galaxies. For the Gronwall-Koo model we estimated the effect of the 
size cut, assuming that the missing galaxies are those with the highest 
redshifts. This procedure gives an indication of the uncertainty in 
$\beta$ due to the size cut. A similar procedure did not change the 
$\beta$'s of the Pozzetti model.

The distributions we used did not contain color information. Therefore we 
do not know the real redshift distributions for the blue and red sample, 
but we will assume that they are comparable.

Another problem is cluster member contamination, which will lower the lensing
signal (or equivalently lower $\beta$). For bright galaxies, the
cluster color-magnitude relation is fairly well defined (cf. 
figure~\ref{colmag}), although there will be a some bright cluster galaxies 
that are blue. At the faint end, the color-magnitude relation becomes 
broader towards the blue, which makes a color selection less effective. 
Cluster member contamination will be worst near the centre of the cluster, 
because there the number density of cluster members is highest and the 
statistics poor. Selecting galaxies based on their colors, allows one to 
lessen the effect of contamination. The effect of cluster member 
contamination on our measurements is discussed in section~7.1.

Gravitational lensing does not only change the shapes of the background
galaxies, but it also magnifies them. As a result the measured flux is 
increased. Due to this magnification bias, the mean redshift of 
galaxies in a certain magnitude bin is a function of distance
to the centre of the cluster (assuming a circular mass distribution) and
therefore also $\beta$ and the critical surface density are functions
of distance to the centre. Though the effect is fairly small for this cluster, 
for more massive clusters $\beta$ may change significantly with radius 
(e.g. Fischer \& Tyson, 1997).

\vbox{
\begin{center}
\begin{tabular}{lcccccc}
\hline
\hline
\multicolumn{1}{l}{name} & \multicolumn{6}{c}{$\beta$} \\
                & GK   & GK1  & LYF  & BF   & POZ & used \\
\hline
bright          & 0.53 & 0.55 & 0.60 & 0.67 & 0.65 & 0.62 \\
faint           & 0.63 & 0.58 & 0.64 & 0.58 & 0.69 & 0.62 \\
bright + faint  & 0.58 & 0.57 & 0.62 & 0.62 & 0.68 & 0.62 \\
blue            & 0.58 & 0.57 & 0.62 & 0.63 & 0.67 & 0.62 \\
red             & 0.55 & 0.55 & 0.61 & 0.65 & 0.66 & 0.62 \\
blue + red      & 0.58 & 0.56 & 0.61 & 0.63 & 0.67 & 0.62 \\
\hline
\hline
\end{tabular}
\end{center}
\begin{small}
{\sc Table~2.} Columns give the mean value for $\beta$ for the
various subsamples using GK the Gronwall-Koo model;
GK1 the Gronwall-Koo model with estimated effect of size
cut at the faint end; LYF photometric redshifts from Lanzetta
et al. (1996); BF Babul-Ferguson model; POZ the Pozzetti model.
In the last column we give the value we use in this paper.
\vspace{0.4cm}
\end{small}}

\vbox{
\begin{center}
\leavevmode
\hbox{%
\epsfxsize=8cm
\epsffile{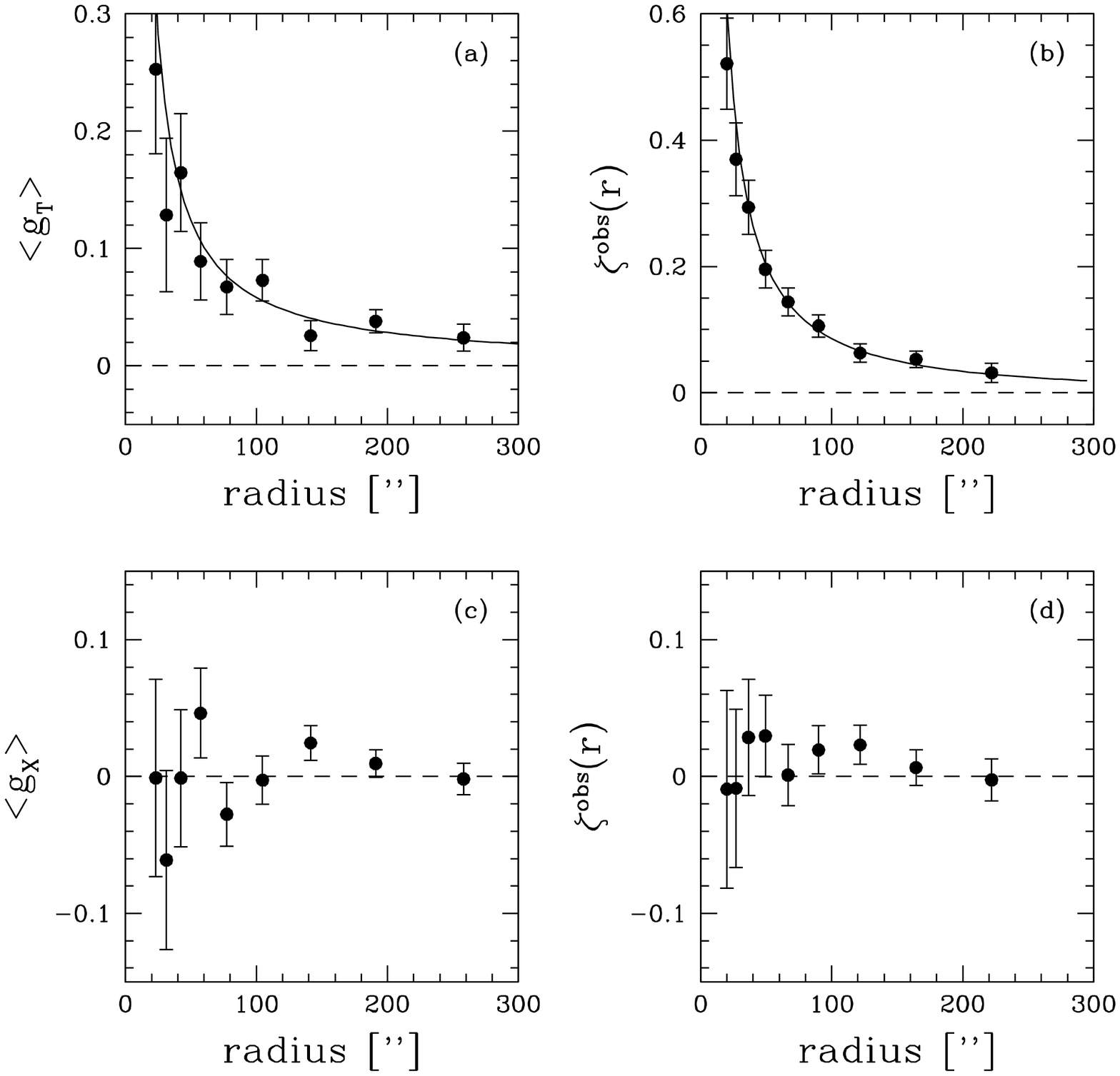}}
\begin{small}
\figcaption{(a) $\langle g_T \rangle$ as a function of radius from the 
centre as measured from our best sample of background galaxies (combined 
sample of blue and red galaxies). (b) The corresponding $\zeta$ profile. 
The solid lines in figures (a) and (b) correspond to the distortion $g_T$ of 
a singular isothermal sphere with a velocity dispersion of 780 km/s, which 
was found by fitting the observations at radii larger than 40 arcseconds. 
(c) The resulting tangential distortion when the phase of the distortion 
is increased by $\pi/2$. The signal should vanish if it is due to 
gravitational lensing. (d) The $\zeta(r)$ corresponding to $g_X$.
\label{bestprof}}
\end{small}
\end{center}}

In Table~2 we list values for $\beta$ for the various subsamples and
redshift distributions. By comparing the various values for $\beta$ 
one gets an idea of the uncertainty in $\beta$. Due to the relatively 
low redshift of Cl~1358, the agreement between the various distributions 
is fairly good. The uncertainty in the mass estimate due to the uncertainty
in the redshifts of the background sources is $\sim 10\%$. We will use a
value of $\beta=0.62$ in this paper. The uncertainty in $\beta$ is 
systematic and will hopefully decrease in the future. We therefore do not 
include the uncertainty in $\beta$ in the error budget of the weak lensing 
analysis.

As the critical surface density depends on the source redshift
distribution, the observed distortion for samples with different 
redshift distributions reflects this difference. Therefore, by
comparing the distortion for different samples of background sources,
one can in principle constrain possible redshift distributions.
We fitted an singular isothermal model ($\kappa(r)=\kappa_0/r$, where
$r$ is in arcsec.) to the observations of the bright and faint sample 
for radii larger than 40 arcsec. For the bright sample we find 
$\kappa_0=5.5\pm1.0$ and for the faint sample $\kappa_0=4.3\pm1.0$, yielding
a ratio of $1.3\pm0.4$ for $\beta_{\rm bright}/\beta_{\rm faint}$. The 
errorbars are calculated from the errors in the tangential distortion. 
This result does not rule out any of the distributions discussed above.

\section{Weak lensing results}

\subsection{Mass estimate}

From the observations one can measure only the distortion $g$, which
is related to the shear $\gamma$ through $g=\gamma/(1-\kappa)$. In the
weak lensing limit ($\kappa \ll 1$) the distortion is equal to
the shear.

To examine the mass distribution of Cl~1358, we first examine the tangential
distortion, $g_T$, which is defined as 
$g_T= -(g_1 \cos~2\phi + g_2 \sin~2\phi$), 
where $\phi$ is the azimuthal angle with respect to the assumed centre 
of the mass distribution. As a measure of the radial surface density profile 
for the cluster we use the statistic (Fahlman et al. 1994; Squires et al. 
1996a)
\begin{equation}
\zeta^{\rm obs}(r,r_{\rm max})=\frac{2}{1-(r/r_{\rm max})^2} 
\int_{r}^{r_{\rm max}} d\ln(r) \langle g_T \rangle
\end{equation}
which gives the mean dimensionless surface density interior to
$r$ relative to the mean in the annulus from $r$ to $r_{\rm max}$ in
the weak lensing limit:
$$\zeta(r,r_{\rm max})=\bar\kappa (r'<r) - \bar\kappa (r<r'<r_{\rm max})$$
The mean tangential distortion $\langle g_T \rangle$ is calculated by 
averaging the tangential distortion in an annulus. $\zeta(r)$ provides a 
lower bound on 
$\bar\kappa (r)$, the mean dimensionless mass surface density within a radius 
$r$, and therefore also on the mass within an aperture of radius $r$. We used 
a value of $300''$  for $r_{max}$. By estimating the mass in the annulus from 
$r$ to $r_{\rm max}$, one can then estimate the mass within an aperture of 
radius $r$.

In Figure~\ref{bestprof}a we show the observed $\langle g_T \rangle$
as a function of radius using the combined catalog of blue and red 
background  galaxies, which we use as our {\it best} catalog. 
In Figure~\ref{bestprof}b the $\zeta$ statistic, which provides a lower 
bound on the mean surface density inside an aperture with radius $r$, is 
plotted. It should be noted here that the points and errorbars in the 
$\zeta(r)$ profile are correlated.

A clear lensing signal is detected although the measurements at
small radii are likely to suffer from cluster member contamination.
Comparing expected counts to the actual counts indicates that approximately
20\% of the galaxies in the innermost bin are cluster galaxies. For the
faint sample, for which we did not a apply a color correction, we estimate 
a fraction of 30\% cluster members in the innermost bin. Furthermore
we found that the software underestimates the distortion for
extremely distorted objects, like the bright arc.

We fit a singular isothermal sphere model to the observations
at radii larger than 40 arcsec to minimize the effect of cluster
member contamination in the innermost region. The distortion equals
the shear divided by $1-\kappa$. In the calculation of the model
distortion we assume that the background galaxies are located 
in a sheet behind the cluster at a redshift that corresponds to $\beta=0.62$.

\vbox{
\begin{center}
\leavevmode
\hbox{%
\epsfxsize=8cm
\epsffile{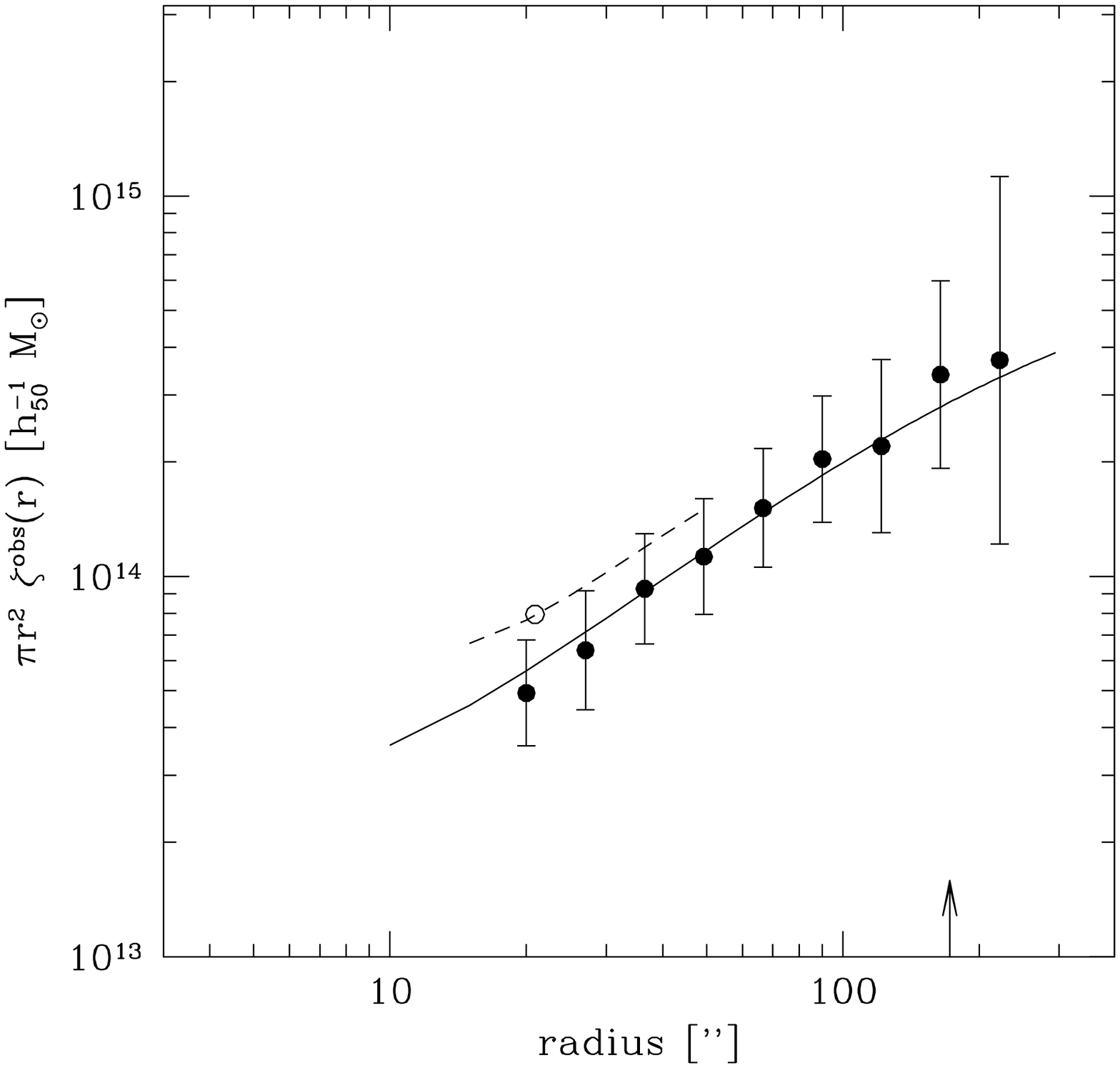}}
\begin{small}
\figcaption{Lower bound on the radial mass profile for our best catalog 
(blue and red galaxies). The errorbars only reflect the uncertainty in 
$\zeta$. The solid line shows the expected profile for our best fit singular 
isothermal sphere which has a velocity dispersion of 780 km/s. The dashed 
line corresponds to the profile calculated from the strong lensing model. 
The open circle gives the enclosed mass at $21''$ for a strong lensing 
model (Franx et al. 1997). The arrow indicates a radius of 1~$h_{50}^{-1}$ 
Mpc.\label{mass}}
\end{small}
\end{center}}

\begin{figure*}
\begin{center}
\leavevmode
\hbox{%
\epsfxsize=16cm
\epsffile[20 155 590 470]{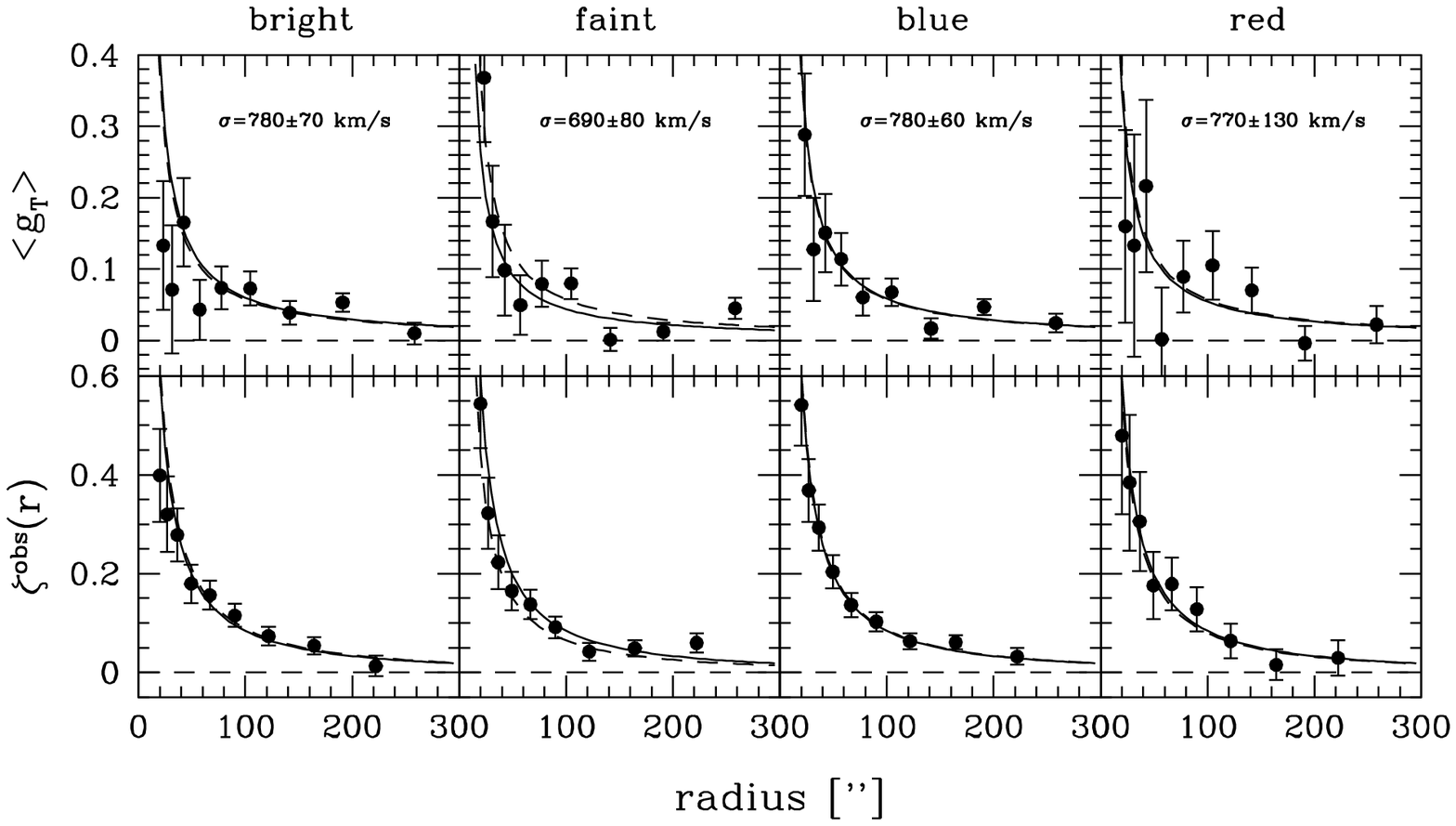}}
\begin{small}
\figcaption{The top row shows the radial tangential distortion profiles
for the bright, the faint, the blue and the red sample. The lower row 
shows the corresponding profiles of the $\zeta(r)$ statistic. The solid 
lines in these figures correspond to the profile of a singular isothermal 
sphere with a velocity dispersion given in Table~3. The 
dashed line corresponds to a velocity dispersion of 780 km/s.\label{profs}}
\end{small}
\end{center}
\end{figure*}

We find a best fitting value for the line-of-sight velocity dispersion of 
$780\pm50$ km/s. The reduced $\chi^2$ of 0.8 indicates that this model fits 
the data well. The uncertainty quoted reflects only the statistical uncertainty 
due to the intrinsic ellipticities of the background galaxies. The systematic 
uncertainty in $\beta$, which introduces an additional 10\% error at most for the 
redshift distributions we considered, is not included in the error budget.

As a consistency check we increased the phase of the distortion 
by $\pi/2$, which is equivalent to rotating all images by 45 degrees.
Formally the surface density corresponding to this distortion is the imaginary
part of the surface density. As the surface density is real the signal from
the rotated distortion should vanish if the observed pattern is due to
gravitational lensing. The measured signal, plotted in the lower row in 
Figure~\ref{bestprof}, is indeed consistent with being zero.

In Figure~\ref{mass} we plot the radial profile 
$\pi r^2 \zeta(r) \Sigma_{\rm crit}$ 
which is a lower bound on the mass M$(<r)$ within an aperture of radius $r$,
using our best catalog. Calculating a similar profile from the strong 
lensing model (Franx et al. 1997) yields the dashed line. The open circle 
in this figure corresponds to the enclosed mass at $21''$ given by the strong 
lensing analysis. The solid line corresponds to the $\zeta(r)$ one would
observe for a singular isothermal sphere model with a velocity dispersion 
of 780 km/s.

From the results of Figure~\ref{mass} we find from this weak lensing analysis 
a lower limit on the mass inside an aperture of radius 1 Mpc of 
$(3.2 \pm 0.9) \times 10^{14}$ M$_\odot$. This mass estimate depends only 
on distortion measurements at radii larger than 1~Mpc. As was expected from 
the tangential distortion measurements, the profile is well fitted by a 
singular isothermal model, with a velocity dispersion of 780 km/s 
(solid line). The projected mass inside 1 Mpc for this model is 
$(4.4 \pm 0.6) \times 10^{14}$ M$_\odot$.

The results from the strong lensing model indicates that the true
$\zeta$-profile differs from the observed one near the centre of the
cluster. However, comparing the singular isothermal model to the 
$\zeta$-profile in the outer region of the cluster indicates that
this model gives a good estimate for the mass within an aperture
of 1~Mpc.

\vbox{
\begin{center}
\begin{tabular}{lcc}
\hline
\hline
name            & $\kappa_0$ [''] & $\sigma$ [km/s] \\
\hline
bright          & 5.5$\pm$1.0 & 780$\pm$70 \\
faint           & 4.3$\pm$1.0 & 690$\pm$80  \\
bright + faint  & 4.9$\pm$0.7 & 740$\pm$50   \\
blue            & 5.4$\pm$0.8 & 780$\pm$60  \\
red             & 5.3$\pm$1.8 & 770$\pm$130 \\
blue + red      & 5.4$\pm$0.7 & 780$\pm$50  \\
\hline
\hline
\end{tabular}
\end{center}
\begin{small}
{\sc Table~3.} 
Results of fitting a singular isothermal sphere model
($\kappa(r)=\kappa_0/r$) to the observed tangential distortion 
$g_T$ at radii larger than 40 arcsec. The second column  gives 
the values found for the model parameter, which corresponds to
half the Einstein radius. In the third column the corresponding 
line-of-sight velocity dispersion is listed, using $\beta$
listed in the last column of table~2
\vspace{0.4cm}
\end{small}}

We also calculated the tangential distortion as a function of radius for 
the various subsamples. The results are shown in the top row of
Figure~\ref{profs}. The corresponding $\zeta(r)$ profiles are shown in the
bottom row. The solid line indicates the profile of the fit of a singular 
isothermal model to the tangential distortion at radii larger than 40 
arcseconds. The corresponding velocity dispersions are presented in 
Table~3. The results for the various subsamples agree well.

The strong lensing mass estimate corresponds to a velocity dispersion of 
970 km/s (Franx et al. 1997), indicating that the weak lensing analysis 
underestimates the mass near the centre of the cluster. Therefore the surface 
density profile is steeper than isothermal near the centre. This is also 
supported by the work of Fisher et al. (1998) who find evidence of substructure
along the line of sight to the centre of the cluster from their 
kinematical analysis. Such small scale substructure along the
line of sight affects the strong lensing analysis more than it
does our weak lensing analysis as we probe the mass distribution
on larger scales. 

The galaxy used in the strong lensing analysis is at a much larger
redshift than the bulk of the galaxies used in the weak lensing
analysis. Therefore a significant cosmological constant can
influence our results. Introducing a cosmological constant,
however, does not decrease the discrepancy between the weak 
and strong lensing.

The weak lensing velocity dispersion is lower than those determined through
dynamical studies. Fisher et al. (1998) find a velocity dispersion
$1027^{+51}_{-45}$ km/s. Their velocity dispersion is biased high due
to substructure along the line of sight near the centre of the cluster.
The cluster was also studied by Carlberg et al. (1997) who find 
$910\pm54$ km/s. The weak lensing estimate agrees marginally with this result.
We find that the mass distribution is flattened (cf. section~7.2). Combined
with the results of Fisher et al. (1998), this indicates that the cluster 
might not be dynamically relaxed.

In many clusters from the CNOC sample arcs are observed (e.g. 
Le F{\`e}vre et al. 1994). For these clusters we estimated the
strong lensing velocity dispersion assuming a singular isothermal
model and that the arc is located at the Einstein radius.
If the redshift of the arc is unknown, we use $z$=1.5 and $z$=3
as a plausible range of source redshifts.
Comparing these results to the velocity dispersion measured by
Carlberg et al. (1997), we find that both estimates agree 
fairly well, although the strong lensing estimate tends to
be slightly larger than the dynamical value.

Smail et al. (1997) find for their sample of clusters that observed
velocity dispersions, estimated from a modest sample of cluster members
($\sim 30$), are typically 50\% higher than inferred from their weak 
lensing analysis. This is quite different from our comparison
using strong lensing for CNOC clusters. The agreement with the
simple strong lensing models indicates that in general the velocity 
dispersions from Carlberg et al. (1997) are representative of the 
mass of the cluster. Also the velocity dispersion inferred from
our weak lensing analysis of Cl~1358 is not that far from the observed
velocity dispersion.

\subsection{Ellipticity of the mass distribution}

As was shown in Schneider \& Bartelmann (1996), one can measure the
quadrupole moments of the mass distribution directly from the observations,
although the expected signal is small. The data we have for Cl~1358
allow us to measure the ellipticity and position angle of the 
mass distribution directly from the observations.

Although the method described in Schneider \& Bartelmann (1996) is
model independent, we assume a mass model to relate the strength 
of the signal to the ellipticity of the mass distribution.
A useful deflection potential $\Psi(r,\phi)$ is
\begin{equation}
\Psi(r,\phi)=2\kappa_0 r \left[{1-q\cos~2(\phi-\alpha)}\right],
\end{equation}
\noindent where $\alpha$ is the position angle. The corresponding 
dimensionless surface density $\kappa$ is given by:
\begin{equation}
\kappa(r,\phi)=\frac{\kappa_0}{r}\left[{1+3q\cos~2(\phi-\alpha)}\right],
\end{equation}
\noindent with an axis ratio $b/a=(1-3q)/(1+3q)$.
For the tangential distortion we can write:
$$g_T(r,\phi)=\langle g_T \rangle
(1+g_{2,c}\cos~2\phi~+~g_{2,s}\sin~2\phi),$$
where $g_{2,c}=3q\cos(2\alpha)$ and $g_{2,s}=3q\sin(2\alpha)$ and 
$\langle g_T \rangle$ is the mean tangential distortion averaged
over a circle. One can measure the axis ratio and position angle by 
measuring $\langle g_T {\rm e}^{2i\phi}\rangle$ on a ring. For this model 
$\langle g_T {\rm e}^{2i\phi}\rangle=\frac{3}{2}q\langle g_T \rangle 
{\rm e}^{2i\alpha}$
As an alternative approach, fitting the model tangential distortion to the 
data, yields the axis ratio and position angle. The advantage of this 
approach is that one does not need data on a complete ring. 

In Figure~\ref{ellips} we present the results of the latter procedure. We find
that both the flattening and the position angle are constant with radius. For
the flattening of the potential we find $q=0.18\pm0.06$, which corresponds
to an axis ratio $b/a=0.30\pm0.15$ for the mass distribution. For the
position angle we find $\alpha=-21^\circ\pm7^\circ$. These results agree
well with the elongation of the light distribution in Figure~\ref{lumdis} 
and the mass distribution in Figure~\ref{recbest}. 

Another model one can use is a singular isothermal ellipsoid
(e.g. Kormann, Schneider, \& Bartelmann 1994). For this model 
we can write:
\begin{equation}
\frac{\langle g_T {\rm e}^{2i\phi} \rangle}{\langle g_T \rangle}=
\frac{2 {\rm E}(1-\frac{a^2}{b^2})-(1+\frac{a^2}{b^2})
{\rm K}(1-\frac{a^2}{b^2})}{(1-\frac{a^2}{b^2}){\rm K}(1-\frac{a^2}{b^2})} 
{\rm e}^{2i\alpha},\\
\end{equation}
\noindent where K($k$) and E($k$) are the complete elliptical integral of 
resp. the first and second kind. Using equation~(7) one can estimate the axis 
ratio. We measured $\langle g_T {\rm e}^{2i\phi}\rangle$ out to 170 
arcseconds from the centre. For this model we found a mean position angle of 
$-18^\circ\pm10^\circ$ and an axis ratio $b/a=0.3^{+0.16}_{-0.12}$, similar 
to what we found previously.

A simple model that reproduces the strong lensing of the $z=4.92$ galaxy is
given by
\begin{equation}
\Psi(x,y)=b\sqrt{r_c^2+x^2(1-\epsilon)+y^2(1+\epsilon)}
\end{equation}
\noindent which corresponds to an isothermal model with core radius and
axis ratio $((1-\epsilon)/(1+\epsilon))^{3/2}$. The best fitting model
parameters are $b=21\farcs05$, $r_c=12\farcs1$, $\epsilon=0.3612$, and
a position angle of $-17^\circ$. The value for $\epsilon$ corresponds
to an axis ratio $b/a=0.32$. Slightly different models give similar results.

The agreement between the weak lensing analysis and the strong 
lensing results is striking, because the strong lensing analysis is most 
sensitive to the mass distribution within the Einstein radius, whereas 
the weak lensing measurements were done at larger radii.

\vbox{
\begin{center}
\leavevmode
\hbox{%
\epsfxsize=8cm
\epsffile{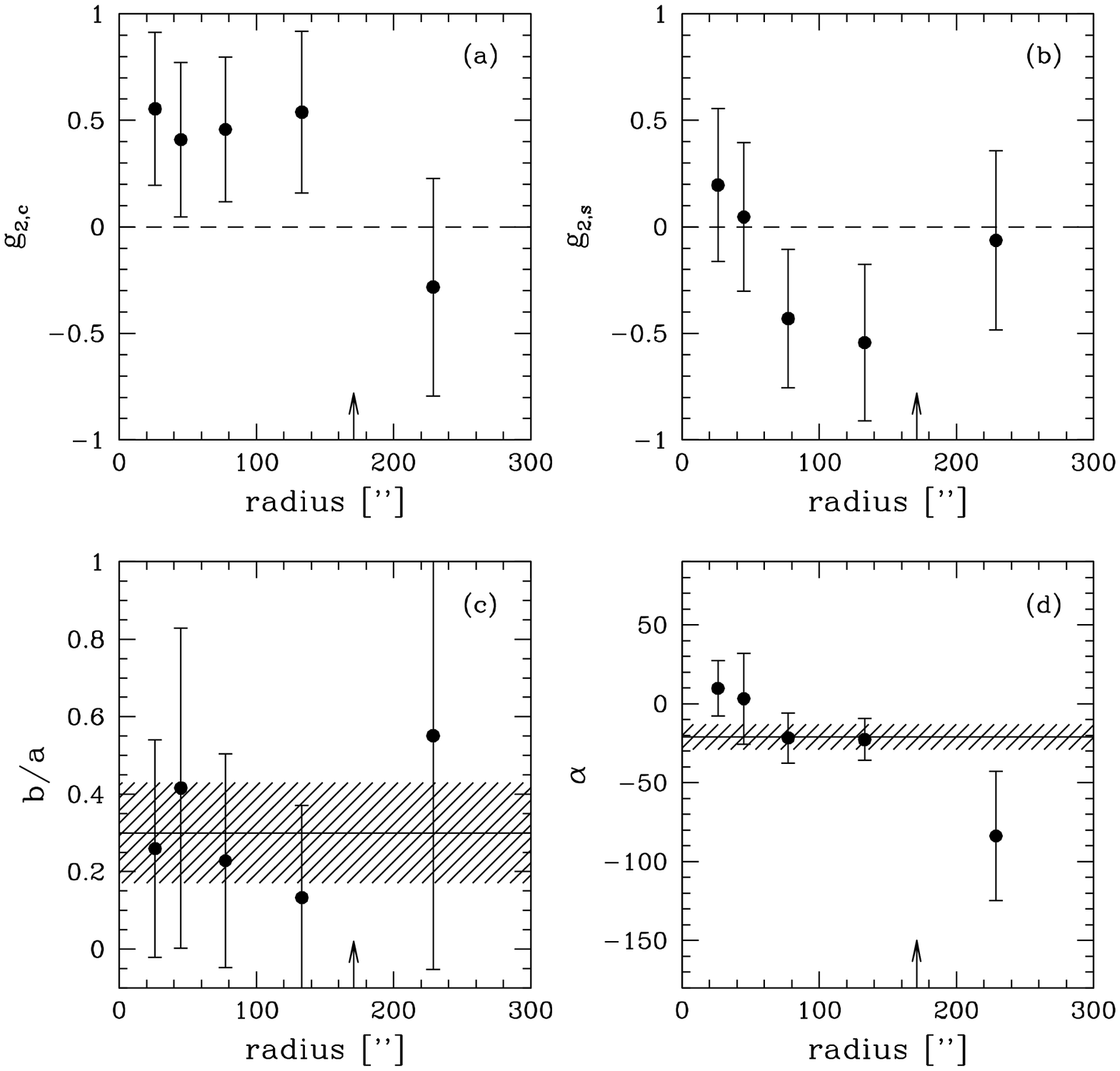}}
\begin{small}
\figcaption{(a) $g_{2,c}=3q\cos(2\alpha)$ as a function of radius and
(b) $g_{2,s}=3q\sin(2\alpha)$ as a function of radius. (c) The
measured axis ratio of the mass distribution $b/a$ versus radius.
We find a mean value of $b/a=0.30\pm0.15$ (indicated by the shaded
region). (d) Position angle $\alpha$ versus radius. We find a mean 
$\alpha=-21^\circ\pm8^\circ$ (indicated by the shaded region).
The arrow indicates a radius of 1~$h_{50}^{-1}$~Mpc.\label{ellips}}
\end{small}
\end{center}}

\subsection{Mass-to-light ratio}

To estimate the mass-to-light ratio, we calculated the expected tangential 
distortion $g_L$ from the observed radial luminosity profile, assuming a
constant mass-to-light ratio. 

For the light distribution we use the sample of bright cluster galaxies 
(cf. section~5). We correct the observed luminosity for the fact that we 
miss approximately 11\% of the light from faint cluster members using 
this catalog  of cluster galaxies, and calculate the distortion $g_L$.
Comparing the observed distortion to $g_L$ allows us to plot the
inferred mass-to-light ratio as a function of radius.

The result, which is shown in Figure~\ref{moverl} is consistent with a 
constant mass-to-light ratio with radius out to $\sim 1h_{50}^{-1}$Mpc. 
Using the procedure described above we find a value of 
$90\pm13~h_{50}{\rm M}_\odot/{\rm L}_{V\odot}$
($95\pm14~h_{50}{\rm M}_\odot/{\rm L}_{V\odot}$ for $q_0=0.1$). As 
was the case for the mass estimate, the error in the mass-to-light ratio 
only reflects the statistical uncertainty due to the random ellipticities 
of the background galaxies. The uncertainty in the redshift distribution
of the background galaxies introduces an extra $10\%$ uncertainty at most
for the redshift distributions considered here.

Another estimate for the mass-to-light ratio is obtained using the 
total mass and light within an aperture of 1~Mpc. Using a total
projected mass of $(4.4\pm 0.6)\times 10^{14}~{\rm M}_\odot$,
which corresponds a singular isothermal sphere with
a velocity dispersion of 780 km/s, and total luminosity of 
$(5.19\pm0.26)\times 10^{12}{\rm L}_{V\odot}$, we find a mass-to-light 
ratio of $85\pm12~h_{50}{\rm M}_\odot/{\rm L}_{V\odot}$.
This second estimate depends on the projected mass within 1~Mpc.
Using the mass-to-light ratio inferred from the light traces mass 
assumption and the observed luminosity one finds a mass of 
$(4.7\pm 0.6)\times 10^{14}~{\rm M}_\odot$ within 1~Mpc, slightly 
higher than we found by fitting a singular isothermal model to the data.

Studies of the fundamental plane for clusters of galaxies at various
redshifts show evidence for galaxy luminosity evolution 
(van Dokkum \& Franx 1996; Kelson et al. 1997). Studying the fundamental 
plane for Cl~1358, Kelson et al. (1997) find that the mass-to-light ratio 
in the $V$ band at the redshift of Cl~1358 is $\sim 30\%$ lower than the 
present day value. Therefore the present day, evolution corrected, 
mass-to-light ratio would be $117\pm17~h_{50}{\rm M}_\odot/{\rm L}_{V\odot}$ 
for Cl~1358.

Cl~1358 is included in the CNOC cluster survey (e.g. Carlberg et al. 1996).
Carlberg, Yee, \& Ellingson (1997) find a value of 
$115\pm15~h_{50}{\rm M}_\odot/{\rm L}_{r \odot}$ for the mass-to-light ratio 
of Cl~1358, using a velocity dispersion of $910\pm54$ km/s and $q_0=0.1$.
Their luminosity is measured in the Gunn $r$-band and is $K$-corrected.

To compare our mass-to-light ratio to the result of Carlberg et al. (1997), 
we transform our mass-to-light ratio to Gunn $r$ and $q_0=0.1$. Using a 
transformation between $B-r$ and $B-V$ from J{\o}rgensen (1994) and the 
$B-V$ colors from the results of van Dokkum et al. (1998) we find that 
our mass-to-light ratio corresponds to a value of 
$84\pm12~h_{50}{\rm M}_\odot/{\rm L}_{r \odot}$. This value, which can be 
compared directly to the value from Carlberg et al. (1997), is
lower than the CNOC value. The difference is due to the fact that our 
weak lensing analysis yields a lower mass than the dynamical analysis.

The range in mass-to-light ratios Carlberg et al. (1996, 1997) find
is fairly small. Carlberg et al. (1996, 1997) therefore conclude 
that the mass-to-light ratios of clusters in the CNOC cluster survey are 
consistent with a universal value. The variance weighted average 
mass-to-light ratio for the CNOC sample is 
$134\pm9~h_{50}{\rm M}_\odot/{\rm L}_{r \odot}$.

For some of these clusters also ground based weak lensing 
analyses are presented in the literature (Fahlman et al. 1994; Squires
et al. 1996b; Smail et al. 1995). Squires et al. (1996b) find
a mass-to-light ratio for Abell~2390 that agrees well with the results 
of Carlberg et al. (1997), whereas Fahlman et al. (1994) find
a much higher mass-to-light ratio for MS~1224+20 from their
weak lensing analysis.

Comparing the velocity dispersions found by Carlberg et al. (1997)
to simple strong lensing estimates (cf. section 7.1) we conclude
that the observed velocity dispersions are in general representative 
for the cluster mass. Due to substructure along the line
of sight, which is an important selection effect for clusters,
both the observed velocity dispersions and strong lensing estimates
might be slightly biased towards higher masses. 

It is therefore difficult to make a firm statement about a universal
mass-to-light ratio by comparing our weak lensing results to the
result of Carlberg et al. (1997). Furthermore, we only measure
the mass-to-light ratio in a fairly small region of the cluster,
whereas the CNOC observations extend to larger radii. Furthermore
luminosity evolution in clusters of galaxies (e.g. van Dokkum \& Franx
1996; Kelson et al. 1997) needs to be taken into account when
comparing clusters at various redshifts.
To examine the issue of a universal mass-to-light ratio, 
more detailed weak lensing measurements of clusters of galaxies are needed.

\vbox{
\begin{center}
\leavevmode
\hbox{%
\epsfxsize=8cm
\epsffile{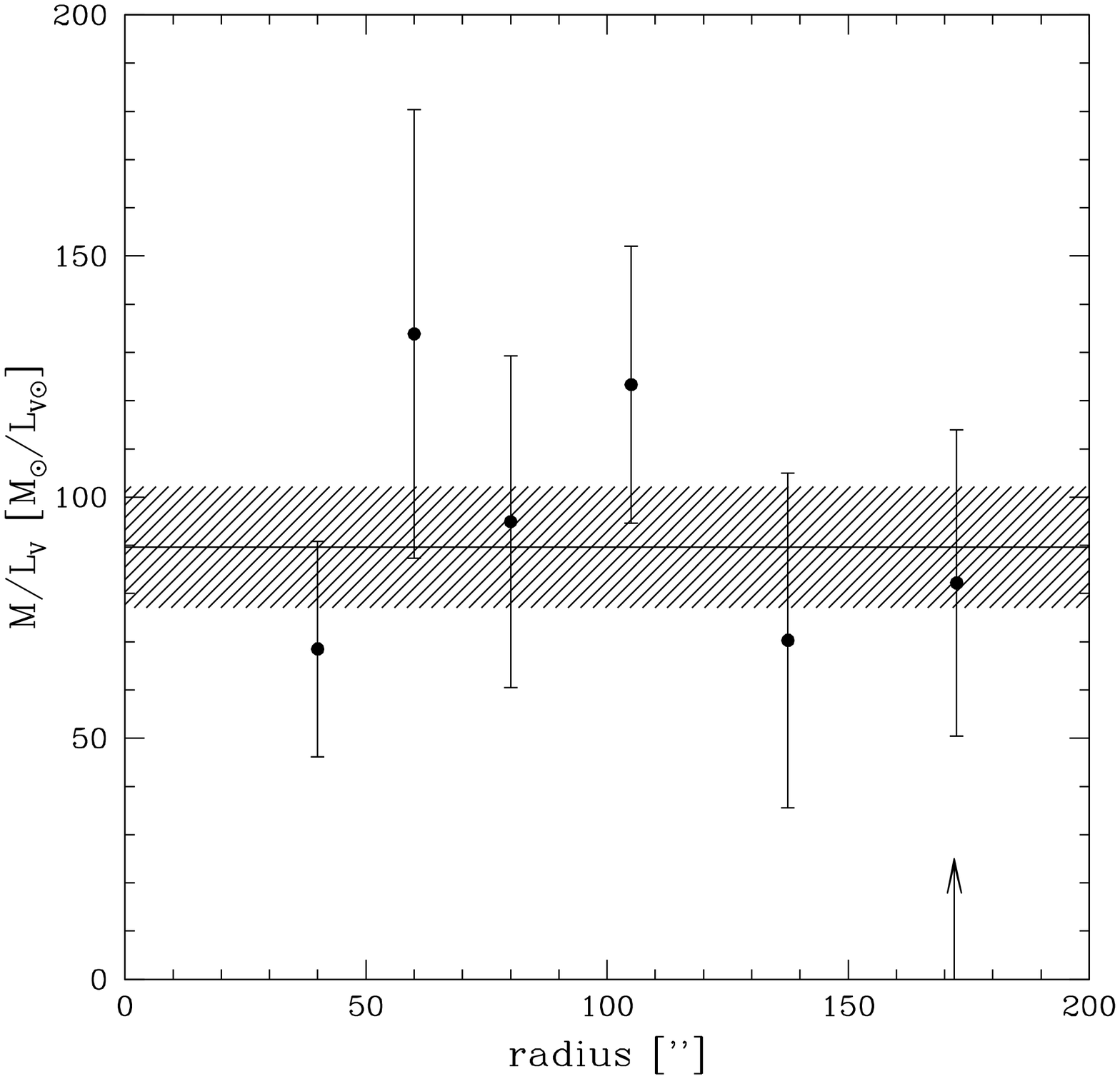}}
\begin{small}
\figcaption{Mass-to-light ratio as a function of radius. From a
sample of bright cluster galaxies we estimated the light content over all 
luminosities using a Schechter luminosity function. The faint end of the 
luminosity function contributes $11\%$ to the total light. We calculate the 
tangential distortion $g_L$ corresponding to the light distribution. Under 
the assumption that the light traces the mass, this allows us to estimate
the mass-to-light ratio as a function of radius. The errorbars only reflect 
the uncertainty in the observations. The uncertainty in the redshift 
distribution of the background galaxies introduces a systematic error
of $10\%$ at most. The solid line corresponds to the mean mass-to-light ratio, 
and the shaded region to the 1$\sigma$ uncertainty. The arrow indicates a 
radius of 1~$h_{50}^{-1}$~Mpc. \label{moverl}}
\end{small}
\end{center}}

\subsection{2-D mass maps}

In the previous sections we have concentrated on the tangential distortion
to estimate the mass and the flattening of the mass distribution of Cl~1358. 
However, from the observed distortion field, one can reconstruct a map of 
the mass surface density up to an unknown additive constant. 

In Figure~\ref{shearmap} we present the observed, smoothed distortion field 
for the best sample of background galaxies (blue and red). We applied a 
Gaussian smoothing with a scale of $0\farcm4$. A clear lensing signal 
centered on the central galaxy is visible.

In Figure~\ref{overlay} the mass map calculated from our best catalog (blue and
red galaxies) is overlayed on the F814W image of Cl~1358. The mass map was 
calculated using the maximum probability extension of the 
original KS algorithm (Kaiser \& Squires 1993; Squires \& Kaiser 1996). 
We used 25 wave modes and a regularization parameter $\alpha=0.05$ and
smoothed the result with a Gaussian of scale 0\farcm4, to allow direct
comparison to the maps of the light distribution.

The peak of the mass map corresponds to the peak in the galaxy light 
distribution. The mass distribution is elongated, roughly in the direction 
indicated by the results of section~7.2, and shows a tail towards the south. 
We carried out bootstrap analyses on the actual mass reconstruction maps. 
It shows that the extension to the south is stable and significant.

\vbox{
\begin{center}
\leavevmode
\hbox{%
\epsfxsize=8cm
\epsffile[70 200 580 700]{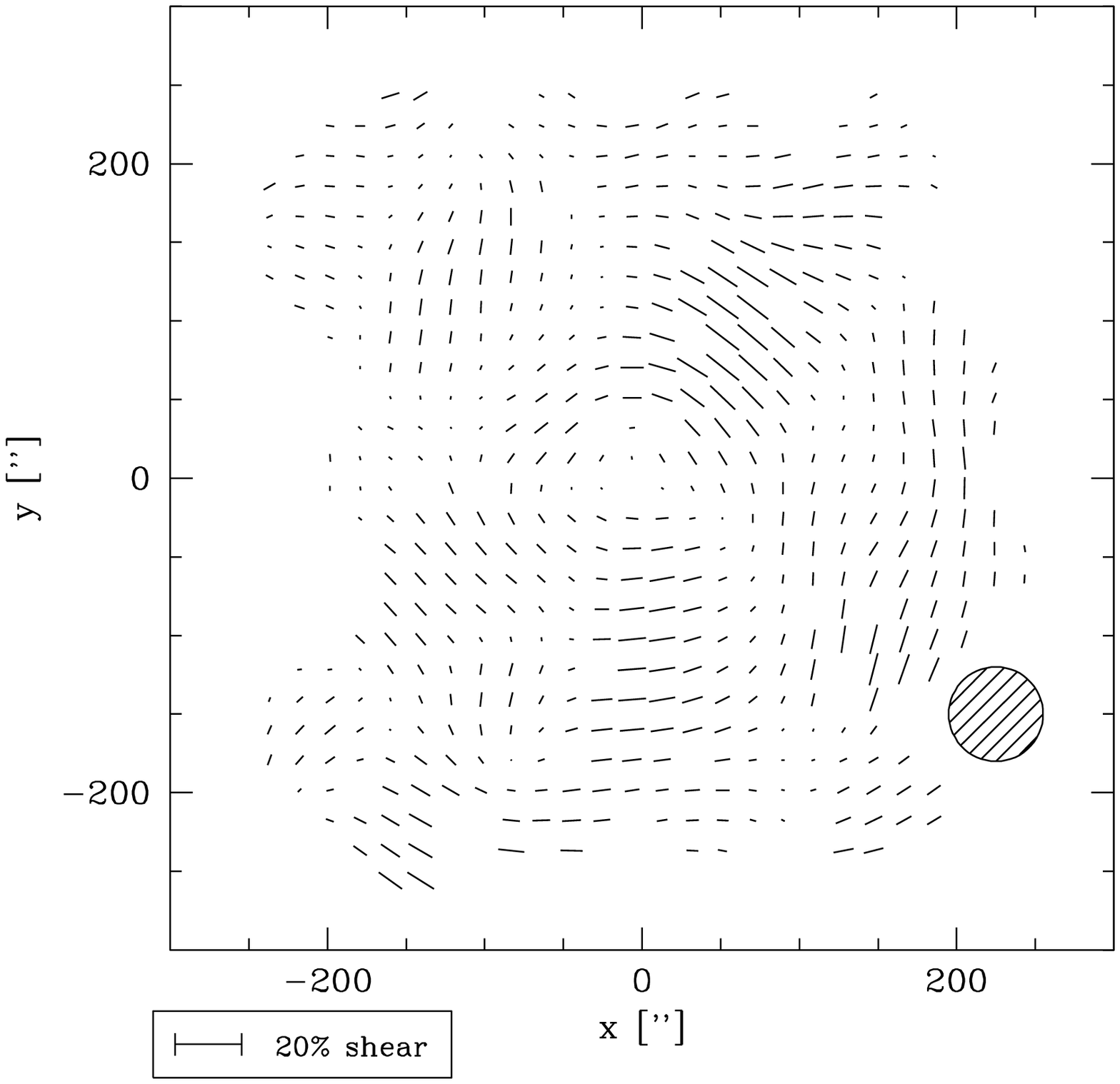}}
\begin{small}
\figcaption{Smoothed distortion field $g$ from the blue and red background 
galaxies. The length of the sticks denote the size of the distortion. The 
distortion field is calculated by smoothing the observed field from the 
individual galaxies with a Gaussian of width $0\farcm4$ (FWHM is indicated 
by the shaded circle). The characteristic pattern due to gravitational 
lensing is clearly visible.\label{shearmap}}
\end{small}
\end{center}}

Comparison with the number density of galaxies, as was presented in
Figure~\ref{lumdis}, shows that we only reproduce the cluster and not 
the enhancements in the number density of background galaxies.
As these concentrations were only visible in the faint galaxy counts,
and were located near the edge of the observed region one would
not expect to recover these concentrations. 

Using the original KS algorithm (Kaiser \& Squires 1993) one can get an 
estimate of the noise in the mass reconstruction. For a map smoothed with
a Gaussian of scale $\theta$, the variance in the dimensionless mass 
surface density is
\begin{equation}
\langle \kappa^2 \rangle = \frac{\langle \gamma^2 \rangle}
{8 \pi \bar n \theta^2}
\end{equation}
\noindent where $\langle \gamma^2 \rangle$ is the variance due to the 
intrinsic shapes of galaxies, for which we found a value of 0.147 from the 
data. For our catalog we have a number density of 42 galaxies arcmin$^{-2}$, 
corresponding to a 1$\sigma$ uncertainty of 0.03 in $\kappa$.
One therefore expects to be able to detect peaks with an associated velocity 
dispersion of $\ge$ 500 km/s at the $\ge 3\sigma$ level over the scale of an 
arcminute (assuming an isothermal model).

We calculated the distortion field corresponding to the mass reconstruction and
compared it to the observations. The residual field is shown in 
Figure~\ref{residual}. The grid shows the layout of the mosaic. At the edges
of the mosaic, the residuals increase due to the lower signal-to-noise, and
in the centre the contribution of cluster members can cause some residuals.

Similar to what was done in Figure~\ref{bestprof}, we calculated the mass map
after increasing the phase of the distortion by $\pi/2$. The result is shown 
in the left panel of Figure~\ref{recbest}. The reconstruction shows a peak in 
the lower left corner and a minimum north of the cluster centre. According to 
equation~9 these would correspond to $\sim 3.6 \sigma$ peaks. Inspection of the
catalog shows that the signal is not due to a local peak but is present on 
scales larger than a WPFC2 chip and therefore cannot be an artefact of the 
camera or the corrections. 

\vbox{
\begin{center}
\leavevmode
\vspace{8.7cm}
\begin{small}
\figcaption{Overlay of the reconstructed mass surface density map from our
best sample of background galaxies overlayed on the F814W image of Cl~1358. 
The mass map was calculated using the maximum probability extension of the
KS algorithm. The mass map was smoothed with a Gaussian of scale 0\farcm4
(The shaded circle indicates the FWHM).
The peak in the projected mass distribution coincides with the central 
elliptical galaxy. The interval between adjacent contour levels is 0.025 
in $\kappa$. According to equation~8, $1\sigma$ corresponds to $\sim 0.03$
in the central region.\label{overlay}}
\end{small}
\end{center}}

Simulations show that the signal-to-noise decreases towards the edges of the
region where we have data. Using the dispersion in $\kappa$ from equation~9,
we find that in $\sim 40\%$ of the simulations '$\ge 4\sigma$' peaks are 
present. These peaks occur generally at the edges of the data region. 
Therefore one cannot trust features in the mass reconstruction that lie 
near the boundaries. The maximum in the lower left corner of 
Figure~\ref{recbest}(a) is due to the low signal-to-noise in this region. 
The minimum is located further away from the boundary. We find from our 
simulations that such minima (or maxima) occur in $\sim 10\%$ of the 
simulations. Only in the centre of the region over which we have data 
does equation~9 hold.

\vbox{
\begin{center}
\leavevmode
\hbox{%
\epsfxsize=8cm
\epsffile[70 200 580 700]{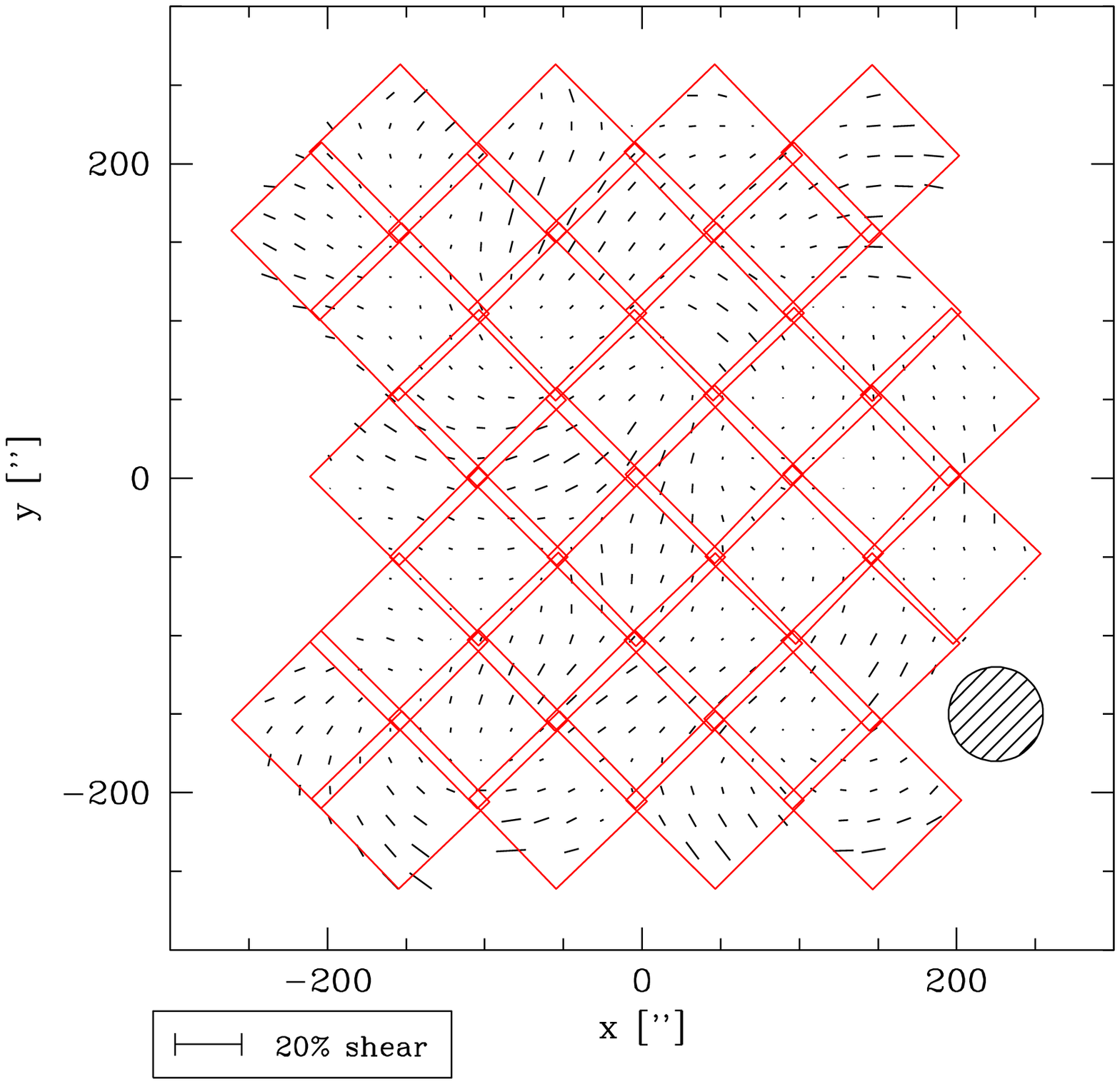}}
\begin{small}
\figcaption{Residual field from the smoothed observed distortion field and the 
distortion calculated from the mass reconstruction. The length of the sticks 
denote the size of the distortion. The shaded circle indicates the FWHM of the 
Gaussian used for the smoothing. The grid shows the layout of the mosaic.
\label{residual}}
\end{small}
\end{center}}

\begin{figure*}
\begin{center}
\vspace{8.5cm}
\begin{small}
\figcaption{(a) The mass surface density reconstruction when the phase of the
distortion is increased by $\pi/2$. This should vanish if the signal
is due to gravitational lensing. (b) Mass reconstruction of the best
catalog (cf. Figure~\protect\ref{overlay}). (c) Result from a finite field 
mass reconstruction algorithm devoloped by Seitz \& Schneider (1996, 1998). 
The interval between adjacent contours is 0.025 in $\kappa$. $1\sigma$ 
corresponds to 0.03 near the centre according to equation~8. All three 
mass reconstructions were smoothed with a Gaussian of scale $0\farcm4$. The 
shaded circles show the FWHM of this smoothing function.\label{recbest}}
\end{small}
\end{center}
\end{figure*}

Although we found a significant elongation of the mass distribution
in section~7.2, which is also seen in the mass reconstruction, we find
little evidence for substructure in the cluster. Cl~1358 was excluded 
from the sample of Carlberg et al. (1997) because of its binary nature,
but the secondary concentration is outside our data region. The results
from Fisher et al. (1997) agree fairly well with the weak lensing mass
map. They found two mass concentrations along the line of sight towards
the centre of the cluster and a third concentration is found south of
the centre, corresponding more or less to our extension to the south.

In the literature much attention has been paid to mass reconstruction
methods (Kaiser \& Squires 1993; Kaiser et al. 1994; Seitz \&
\& Schneider 1996; Schneider 1995; Squires \& Kaiser 1996). Although
we use the maximum probability extension of the original Kaiser \&
Squires (1993) algorithm we also calculated a finite field mass
map using the method described in Seitz \& Schneider (1996, 1998). 
The result is shown in Figure~\ref{recbest} (right panel). We smoothed the 
distortion field with the same Gaussian that was used to smooth the mass map 
from the maximum probability method. The two reconstructions are very similar.

\subsection{Comparison to X-ray results}

The X-ray emission of Cl~1358 was studied by Bautz et al. (1997),
using ROSAT and ASCA data. They show that a substantial fraction
($10^{44}~{\rm erg s}^{-1}$) of the total X-ray luminosity of 
$L_x~(0.2-4.5)~{\rm keV}~=~7 \times 10^{44} h_{50}^{-2}~{\rm erg/s}$
is emitted from cool gas, indicating that a cooling flow exists in
this cluster. The surface brightness distribution, with PSPC
resolution, is unimodal and smooth. 

Ignoring the influence of the cooling flow on the integrated ASCA spectrum 
(but accounting for it in the image analysis,) and making all 
the standard assumptions, the ROSAT surface brightness profile and the 
ASCA temperature give a preliminary mass of 
$6.6\times 10^{14}~h_{50}^{-1}~{\rm M}_\odot$ within 1 Mpc radius 
(Bautz 1997, priv. comm.), although the mass uncertainty is not less 
than 30\%. 

Allen (1997) analysed the ROSAT and ASCA data on Cl~1358, taking
into account the effects of cooling flows on the X-ray images
and spectra. The results of Allen (1997) imply a projected mass of
$4.2^{+4.1}_{-0.8}\times 10^{14}~{\rm M}_\odot$ (90\% confidence
limits) within an aperture of 1 Mpc from the cluster centre,
in excellent agreement with the weak lensing mass estimate of 
$(4.4\pm 0.6)\times 10^{14}~{\rm M}_\odot$, which assumed a
singular isothermal sphere model with a velocity dispersion of
780 km/s.

The gas mass estimated by Bautz (priv. comm.) is 
$6.4\times 10^{13}{\rm M}_\odot$, with a relative error of at least 
$15\%$. Combining the mass of the gas and our weak lensing estimate of 
the total mass yield a lower limit of $f_b >0.15~h_{50}^{-3/2}$ for the 
baryon fraction which agrees well with the results of White, Jones, \& 
Forman (1997).

\subsection{Magnification bias}

Gravitational lensing does not only distort the images of background 
galaxies, it also magnifies them (e.g. Broadhurst, Taylor, \& Peacock 1995; 
Mellier et al. 1996). One consequence of this is that it changes the local
number density of background galaxies. Another way of detecting
the magnification bias involves measuring the sizes of objects
(Bartelmann \& Narayan 1995).

The observable number density of background galaxies depends on the relative
strength of a deflection effect with respect to a magnification effect.
The lens tends to deflect sources away from the centre, thus inducing a 
lower number density, whereas the magnification effect makes the objects
larger and thus increases the detectable number of objects 
(Broadhurst et al. 1995; Fort et al 1996; Taylor et al. 1998).

The expected number density of background galaxies is
$$ N(r)=N_0 \mu(r)^{2.5\alpha-1},$$
where $\mu$ is the magnification at radius $r$ and 
$\alpha$ is the slope of the galaxy counts ${\rm d}\log(N)/{\rm d}m$.

It is found that the count slopes for red objects are lower than
those found for blue objects (Broadhurst 1995). Such an effect would result 
in a decrease of the ratio of the number density of red over blue objects 
towards the centre of the cluster. We used our catalog of objects detected 
both in F606W and F814W. The red sample is created using all objects having 
F606W$-$F814W$>1.4$ and $24<$F814W$<26.5$.
The blue sample is defined by objects with F606W$-$F814W$<0.8$ and 
$24<$F814W$<26.5$. This large separation in color gives a minimal contribution
from cluster galaxies. The count slopes $\alpha_{\rm red}\approx 0.25$ and
$\alpha_{blue}\approx 0.45$ indicate that one could measure the
magnification, given a signal that is strong enough. 

In Figure~\ref{magbias} we plot the ratio of number densities of red and blue 
objects as a function of radius. The solid line shown in this figure shows 
the expected signal for a singular isothermal sphere with a velocity 
dispersion of 780 km/s. This already shows that the expected signal is 
small. Given the fairly low expected signal-to-noise for this method , 
our measurements are consistent with a constant ratio, i.e. consistent 
with no detectable magnification. 

This method to detect a magnification bias, however, suffers from
several problems. The results are very sensitive to clustering of
the background objects. A local lack of red objects could either be
due to the bias or just to a real underdensity of red
objects. Furthermore the Poisson noise on the counts is fairly large
compared to the expected signal. 

\vbox{
\begin{center}
\leavevmode
\hbox{%
\epsfxsize=8cm
\epsffile[70 200 580 700]{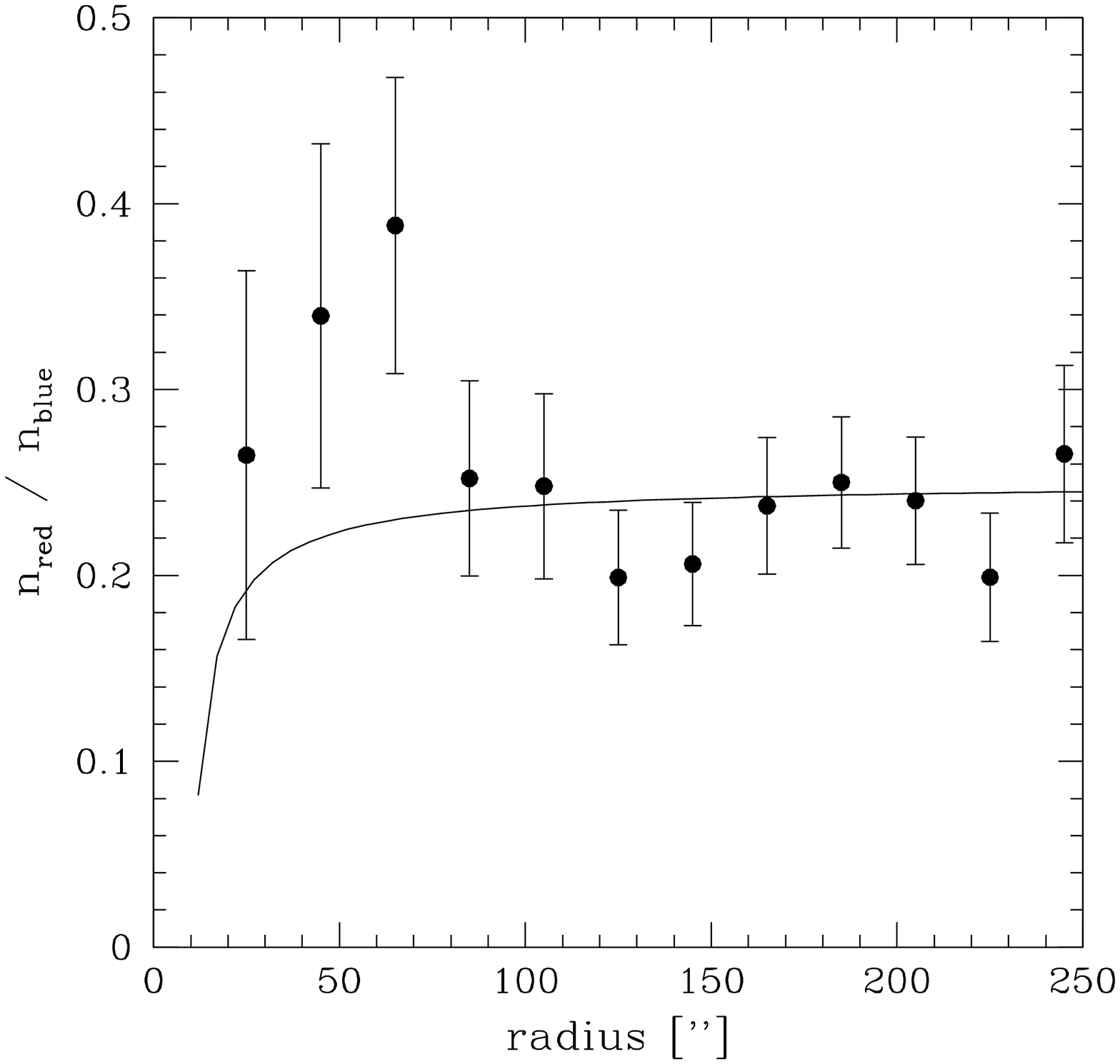}}
\begin{small}
\figcaption{Ratio of the number density of red objects and blue objects as a
function of radius from the cluster centre. The plotted points are
independent measurements of this ratio. Red objects have a F606W$-$F814W color
$\ge 1.4$ and blue objects have F606W$-$F814W$\le 0.8$. The solid line 
indicates the expected signal, for a singular isothermal sphere with a 
velocity dispersion of 780 km/s. The measurements are consistent with a 
constant ratio.\label{magbias}}
\end{small}
\end{center}}

An argument for measuring the magnification bias from number counts
is that one does not require extremely good seeing as one just counts
objects. The data presented in the paper here yields high number densities
of background galaxies. Comparing the null-detection of the magnification
bias to the results of the shear measurements, shows that measuring 
the magnification bias through number counts is hard for most clusters of
galaxies. For very massive clusters, however, the magnification
signal might be detected (Taylor et al. 1998)

\section{Conclusions}

We have mapped the light and mass distributions of the cluster of 
galaxies Cl~1358+62, using a large mosaic of HST observations in F606W and
F814W. The total field of view of these observations is approximately 8 by 8 
arcminutes. This provides us with an ideal dataset for a weak lensing analysis 
of the cluster.

The WFPC2 point spread function is highly anisotropic at the edges of the 
individual chips. This systematically distorts the shapes of the faint
background galaxies. We improved the correction technique developed by KSB95
and LK97 and used it to correct the shapes for PSF effects. We exclude galaxies
with a scale length less than 1.2 pixels, because simulations show that these
object cannot be corrected well. We show that this scheme allows a
straightforward correction for shear introduced by camera distortion.

We detect 4175 objects with scale lengths larger than 1.2 pixels, corresponding
to a number density of 79 galaxies arcmin$^{-2}$. From this catalog we select a
sample of blue and red background galaxies, which contains 2228 galaxies
(42 galaxies arcmin$^{-2}$). 

We detect a weak lensing signal out to $\sim 1.5$~Mpc from the cluster 
centre. Fitting a singular isothermal sphere model to the observed 
tangential distortion shows that the data are consistent with a singular
isothermal sphere with a velocity dispersion of $780\pm50$ km/s.
For this model, the corresponding projected mass within a radius of 
1~Mpc from the cluster centre is $(4.4\pm0.6) \times 10^{14}\rm{M}_\odot$.
The errors quoted here only include the statistical uncertainty due
to the intrinsic ellipticities of the background galaxies.
The uncertain redshift distribution of these background sources
introduces an additional systematic error of at most 10\% for 
the redshift distributions considered here.

The weak lensing velocity dispersion is lower than the dynamical estimates of
$1027^{+51}_{-45}$ km/s (Fisher et al. 1997) and $910\pm54$ km/s 
(Carlberg et al. 1996). A strong lensing model with a velocity dispersion 
of 970 km/s (Franx et al. 1997), indicates that the weak lensing analysis 
underestimates the mass in the central region and that the projected mass 
profile is steeper than isothermal in the centre.

Comparing the observed velocity dispersions of cluster in the 
CNOC sample (Carlberg et al. 1997) to simple strong lensing
estimates we find no evidence that the observed velocity
dispersions are much higher than those inferred from lensing
(Smail et al. 1997). Therefore the velocity dispersions
measured by Carlberg et al. (1997) are in general representative
for the cluster masses.

A kinematical study of Cl~1358 by Fisher et al. (1997) shows that there
are two mass concentrations along the line of sight, of which one
corresponds to the BCG and its companions. This elongation, as well
as the flattening of the projected mass distribution, can increase
the observed velocity dispersion.

The projected mass within 1 Mpc inferred from X-ray observations is
$(4.2^{+4.1}_{-0.8} \times 10^{14}\rm{M}_\odot$ (90\% confidence) (Allen 1997),
in good agreement with the weak lensing mass.
Bautz (priv. comm.) estimated the total gas mass to be 
$(6.4\pm1.0)\times 10^{13}{\rm M}_\odot$. Combining this result with the
weak lensing mass yields a lower limit of $f_b>0.15~h_{50}^{-3/2}$ for
the baryon fraction, in good agreement with White et al. (1997).

The elongation of the mass distribution is measured directly from the data.
We find that the mass distribution is elongated with a position
angle of $-21^\circ\pm7^\circ$ measured from the north. Assuming 
an ellipsoidal mass distribution we find an axis ratio of 
$0.3\pm0.15$. This result agrees well with what is found from the light 
distribution, and the mass reconstruction.

A strong lensing analysis based on the red arc (Franx et al. 1997), gives 
nearly the same results: a position angle of $-17^\circ$ and an axis ratio 
$b/a=0.32$. As the weak and strong lensing analyses are probing different 
regions in the cluster, the results indicate that the flattening and position 
angle of the mass distribution are fairly constant with radius.

To map the total projected mass distribution, we used the maximum probability
extension of the original KS algorithm (Kaiser \& Squires 1993; 
Squires \& Kaiser 1996). We compared the resulting mass map to the result 
from a finite field reconstruction algorithm developed by Seitz 
\& Schneider (1996, 1998). The two mass maps are very similar.

The peak of the reconstructed mass distribution coincides with the central
elliptical galaxy and the peak of the light distribution. Also the overall
shape resembles that of the light. The reconstruction shows an extension to
the south, as is seen in the optical light. 

Assuming a constant mass-to-light ratio, we calculated the 
tangential distortion corresponding to the observed radial light
profile. Comparing the result to the observed distortion yields
the mass-to-light ratio, provided that the light traces the mass. We 
find that the mass-to-light ratio is consistent with being constant with 
radius and estimate a value of 
$(90\pm 13)~h_{50}~{\rm M}_\odot/{\rm L}_{\odot V}$.

The corresponding mass-to-light ratio in Gunn $r$ (and $q_0=0.1$) is
$(84\pm 12)~h_{50}~{\rm M}_\odot/{\rm L}_{\odot r}$ is lower
than the value of $(115\pm 15)~h_{50}~{\rm M}_\odot/{\rm L}_{\odot r}$ 
measured by Carlberg et al. (1997), due to the fact we derive a lower
mass for Cl~1358. 

Our mass-to-light ratio is significantly lower than the average 
mass-to-light ratio found from the sample of clusters in the 
CNOC cluster survey (Carlberg et al. 1996; Carlberg et al. 1997).
From the results presented in this paper it is, however, not
possible to make a firm statement about the universality of 
cluster mass-to-light ratios. Due to substructure along the
line of sight, the CNOC results might be biases somewhat
high. Furthermore we only probe a fairly small region of 
the cluster compared to Carlberg et al. (1997). Comparing
clusters at various redshifts, luminosity evolution needs
to be taken into account (e.g. van Dokkum \& Franx 1996;
Kelson et al. 1997). To investigate the issue of a
universal mass-to-light ratio in more detail, more detailed 
weak lensing analyses of clusters of galaxies are needed.

The analysis presented in this paper shows that mosaics of HST observations
provide an excellent opportunity to do detailed weak lensing studies 
of clusters of galaxies. The high number density of background objects, 
combined with the small corrections for the size of the point spread function 
allow us to calculate high resolution mass maps, and precise mass estimates.

\acknowledgments

The authors would like to thank Nick Kaiser for making the
imcat software available. It is a pleasure to thank Peter
Schneider for useful discussion and for making available
his mass reconstruction method. We also thank Caryl Gronwall, 
Rychard Bouwens, Joseph Silk, and Harry Ferguson for providing their $n(m,z)$ 
distributions and Mark Bautz for sharing his information on the X-ray 
observations. The initial data reduction, done by Pieter van Dokkum, is 
obviously invaluable for this analysis. The comments from the referee, 
Dr. Gary Bernstein, helped to clarify the paper significantly.

\appendix
\section{Correcting for the PSF and camera distortion}

The point spread function and the camera distortions change the shapes
of the galaxies needed for the weak lensing analysis. The effect of the
PSF is a smearing and depends on the size of the object. A camera distortion
introduces a genuine shear, which is independent of the size of the object, but
depends on its shape. One must correct the shapes of the galaxies for these
effects.

\subsection{Correcting for the PSF}

To estimate the shear from our observations, we followed the method described 
in Kaiser, Squires \& Broadhurst (1995, KSB95) and Luppino \& Kaiser (1997, 
LK97). The HST observations differ from usual ground based observations, in 
the sense that the HST point spread function is badly sampled and that the 
PSF is anisotropic at large radii. It is therefore not obvious that the 
correction scheme described in KSB95 can be applied to HST data. We 
examined the derivation presented in KSB95 and tested the method by 
simulations.

To quantify the distortion, we combine the quadrupole moments to form a two
component polarization (Blandford et al. 1991):

\begin{equation}
e_1 = \frac{I_{11}-I_{22}}{I_{11}+I_{22}}~{\rm and}
~e_2 = \frac{2I_{12}}{I_{11}+I_{22}}.
\end{equation}

To suppress divergent sky noise contributions in the measurements of the
quadrupole moments we use a Gaussian weight function. Thus the quadrupole
moments are evaluated using:

\begin{equation} 
I_{ij} = \int d^2\theta W(\theta) \theta_i \theta_j f(\theta).
\end{equation}

\noindent Here $(\theta_1, \theta_2) = (0,0)$ is the assumed centre 
of the object. The centre is choosen to be the point where the
weighted dipole moments are zero. We use a Gaussian with a dispersion 
that is proportional to $r_g$, the radius of maximum significance (this radius 
is given by the peak finder, KSB95). For such a weighting scheme the 
first order shift in polarization due to the shear is given by:

\begin{equation}
\delta e_\alpha=P^{\rm sh}_{\alpha \beta} \gamma_\beta,
\end{equation}
\noindent where the shear polarizability $P^{\rm sh}_{\alpha \beta}$ is given 
by
\begin{equation}
P^{\rm sh}_{\alpha \beta}=X^{\rm sh}_{\alpha \beta} - e_\alpha e^{\rm sh}_\beta
\end{equation}
\noindent and where $X^{\rm sh}_{\alpha \beta}$ and $e^{\rm sh}_\alpha$ are 
defined as
\begin{equation}
X^{\rm sh}_{\alpha \beta}=\frac{1}{I_{11}+I_{22}}\int d^2\theta
\left[\begin{array}{cc}
2W\theta^2+2W'(\theta_1^2-\theta_2^2)^2    & 4W'(\theta_1^2-\theta_2^2)
\theta_1\theta_2 \\
4W'(\theta_1^2-\theta_2^2)\theta_1\theta_2 & 2W\theta^2 + 8 W'\theta_1^2
\theta_2^2\\
\end{array}\right]f(\vec\theta)
\end{equation}
\noindent and
\begin{equation}
e^{\rm sh}_\alpha= 2 e_\alpha + \frac{2}{I_{11}+I_{22}}\int d^2\theta 
\left(\begin{array}{c}
\theta_1^2-\theta_2^2\\
2\theta_1\theta_2
\end{array}\right) W'\theta^2 f(\vec\theta),
\end{equation}

\noindent where $W'$ denotes the derivative of $W$ with respect to $\theta^2$.
The shear polarizability  gives the response of an object to an applied 
shear and can be calculated from the data. Note, however, that equation~A6
differs from B12 in KSB95. This difference is significant for elongated
objects. Also the effect of seeing is not included in this estimate of the 
shear polarizability. LK97 present a nice way to solve this problem.

An anisotropic point spread function changes the shapes of observed objects. 
Before applying equation~A3, one must therefore correct the measured 
polarizations for this effect. KSB95 give a correction if the PSF anisotropy 
can be written as a convolution of a small, normalized, anisotropic PSF with a
circular seeing disc. Such a point spread function changes the quadrupole 
moments according to

\begin{equation}
I_{ij}'=I_{ij} + q_{lm}Z_{lmij},
\end{equation}
\noindent where $q_{lm}$ are the unweighted quadrupole moments of the PSF and
where $Z_{lmij}$ is given by
\begin{equation}
Z_{lmij}=\frac{1}{2}\int d^2\theta f(\vec\theta)
\frac{\partial^2[W(\vec\theta)\theta_i \theta_j]}
{\partial \theta_l \partial \theta_m}.
\end{equation}
\noindent This equation is similar to equation~A4 in KSB95 except for the 
factor $\frac{1}{2}$. The change in polarization due to an anisotropy in the 
point spread function is given by
\begin{equation}
\delta e_\alpha=P^{\rm sm}_{\alpha \beta} p_\beta,
\end{equation}
\noindent where $p_\alpha\equiv(q_{11}-q_{22},~2q_{12})$ is a 
measure of the PSF anisotropy and where $P^{\rm sm}_{\alpha \beta}$ 
is the smear polarizability, defined as
\begin{equation}
P^{\rm sm}_{\alpha \beta}=X^{\rm sm}_{\alpha \beta} - e_\alpha e^{\rm sm}_\beta
\end{equation}
\noindent and where $X^{\rm sm}_{\alpha \beta}$ and $e^{\rm sm}_\alpha$ are 
given by
\begin{equation}
X^{\rm sm}_{\alpha \beta}=\frac{1}{I_{11}+I_{22}}\int d^2\theta
\left[\begin{array}{cc}
W+2W' \theta^2 + W''(\theta_1^2-\theta_2^2)^2 & 2W''(\theta_1^2-\theta_2^2)
\theta_1\theta_2 \\
2W''(\theta_1^2-\theta_2^2)\theta_1\theta_2 & W+2W' \theta^2 + 4W''\theta_1^2
\theta_2^2  \\
\end{array}\right]f(\vec\theta)
\end{equation}
\noindent and
\begin{equation}
e^{\rm sm}_\alpha= \frac{1}{I_{11}+I_{22}}\int d^2\theta 
\left(\begin{array}{c}
\theta_1^2-\theta_2^2\\
2\theta_1\theta_2
\end{array}\right) (2W'+W''\theta^2)f(\vec\theta),
\end{equation}
\noindent where prime denotes differentiation with respect to $\theta^2$.

Apart from a factor of a half, equation~A12 is slightly different from 
equation~A12 in KSB95. The effect of this difference is fairly small 
and was therefore not noticed in previous simulations. These differences 
were also noticed independently by Hannelore H\"ammerle.

According to equation~A7, one should use the unweighted quadrupole moments of 
the stars to quantify the anisotropy of the PSF. We use a different estimate

\begin{equation}
p_\alpha=\frac{e_\alpha^*}{P^{\rm sm*}_{\alpha\alpha}}(r_g),
\end{equation}

\noindent where the symbol $*$ indicates that the parameters are calculated 
from the stars. We use the same weight function that was used to calculate 
the parameters of the object we want to correct. In KSB95 a similar estimate 
for $p_{\alpha}$ is used, although there the weight function used to 
calculate the PSF parameters is not matched to the object one wants to correct.

As is demonstrated below, the original KSB95 scheme cannot be applied to more 
complex point spread functions. As mentioned above the correction holds if the
PSF can be described as a convolution of a post-seeing circularly smeared 
image with a small anisotropic PSF. The HST point spread function cannot be 
described this way. Oversampled point spread functions calculated by Tiny Tim 
(Krist \& Hook 1996) show that the anisotropy is mainly caused by the 
anisotropy of the diffraction rings.

As a compromise, we adapt the weight function when calculating PSF parameters.
In appendices B, C and D our choice is justified. Although it is not a perfect
scheme, the simulations presented in the appendices below show that our
approach works fairly well.

Having corrected for the PSF anisotropy, one still needs to calculate the 
shear from the measured polarizations. The calculated shear polarizability
gives a good estimate in the absence of seeing. The polarizations of small 
objects, however, are decreased because the PSF circularizes objects. Up to 
now people used HST images to estimate
the correction factor for their ground based observations (KSB95).
A similar scheme is not at hand when using HST data.

Fortunately, LK97 present a very useful way to calculate a 'pre-seeing' shear 
polarizability
\begin{equation}
P^\gamma=P^{\rm sh}-\frac{P^{\rm sh}_{*}}{P^{\rm sm}_{*}}P^{\rm sm}
\end{equation}
\noindent where the subscript $*$ denotes once more values measured from the 
stars. This 'pre-seeing' shear polarizability can be calculated directly 
from the observations.

In LK97 nothing is mentioned about the weighting one should use. The weight
function that is used in the analysis of the galaxies is a Gaussian with
a scale length $r_g$, equal to the value given by the peakfinder. But in 
principle one might take any weight function. 

This issue is important because in general $P^{\rm sh}_{*}/P^{\rm sm}_{*}$ 
will be different for different $r_g$. An intuitive guess is that 
$P^{\rm sh}_{*}$ and $P^{\rm sm}_{*}$ have to be calculated from the star 
using the same weight function as was used in calculating the 
polarizabilities for the object one wants to correct, similar to what we 
found for the correction for the PSF anisotropy. In appendices~C and D this 
is investigated in more detail.

\subsection{Correcting the camera distortion}

The shear introduced by camera distortion is usually small. However, using
the results of the previous section, one can easily correct for this effect.
The derivation is similar to that leading to the 'pre-seeing' shear 
polarizability in LK97.

We measure the polarization and polarizabilities of stars and galaxies from
our observations. In the previous section we did not take the effect of a
camera shear into account. Such a shear $\delta$ changes the shape of a star:
\begin{equation}
e^{\rm obs}_*=e_*+P^{\rm sh}_* \delta,
\end{equation}
\noindent where $e_*$ is the polarization of the star in the absence of a
camera distortion and $e_*^{\rm obs}$ the observed polarization. For a 
galaxy we measure a polarization
\begin{equation}
e^{\rm obs}=e^{\rm true}+P^\gamma \gamma + \frac{P^{\rm sm}}
{P^{\rm sm}_*}e^{\rm obs}_* + P^{\rm sh}\delta,
\end{equation}
with $e^{\rm true}$ the intrinsic polarization of the galaxy and $\gamma$ the 
shear due to gravitational lensing. Using equation~A15 we get
\begin{equation}
e^{\rm obs}=e^{\rm true}+P^\gamma \gamma + P^{\rm sm}
\frac{e_*}{P^{\rm sm}_*} - P^{\rm sh} \frac{P^{\rm sh}_*}{P^{\rm sm}_*} \delta
+ P^{\rm sh}\delta,
\end{equation}
To correct for the PSF anisotropy, we use the observed polarization of the 
stars in equation~2. Thus we find that
\begin{equation}
e^{\rm obs}=e^{\rm true} + P^{\rm sm} p + P^\gamma (\gamma + \delta),
\end{equation}
\noindent where $p$ is $e^{\rm obs}_*/P^{\rm sm}_*$. 

After correcting for the PSF anisotropy, one finds that the observed shear is 
equal to $\gamma+\delta$. To correct for the camera distortion, one just has to
subtract $\delta$ from the observed shear field.

\twocolumn

\section{Anisotropy correction}

In KSB95 the authors quantify the anisotropy of the PSF, $p_\alpha$, by dividing 
the polarization of a star by its smear polarizability.  They implicitly assume 
that the value of $p_\alpha$ thus derived does not depend on the weight function;
formally it corresponds to the unweighted quadrupole moments of the PSF.

The HST point spread function is fairly complicated, and therefore
we tested several correction schemes for the PSF anisotropy:
(i) $p_\alpha$ calculated with a weight function of size matched to 
the stellar image, as in KSB95, (ii) using the true $p_\alpha$ 
(unweighted), and (iii) using a weight function of size matched
to the galaxy image to be corrected.

\vbox{
\begin{center}
\leavevmode
\hbox{%
\epsfxsize=8cm
\epsffile[20 160 580 650]{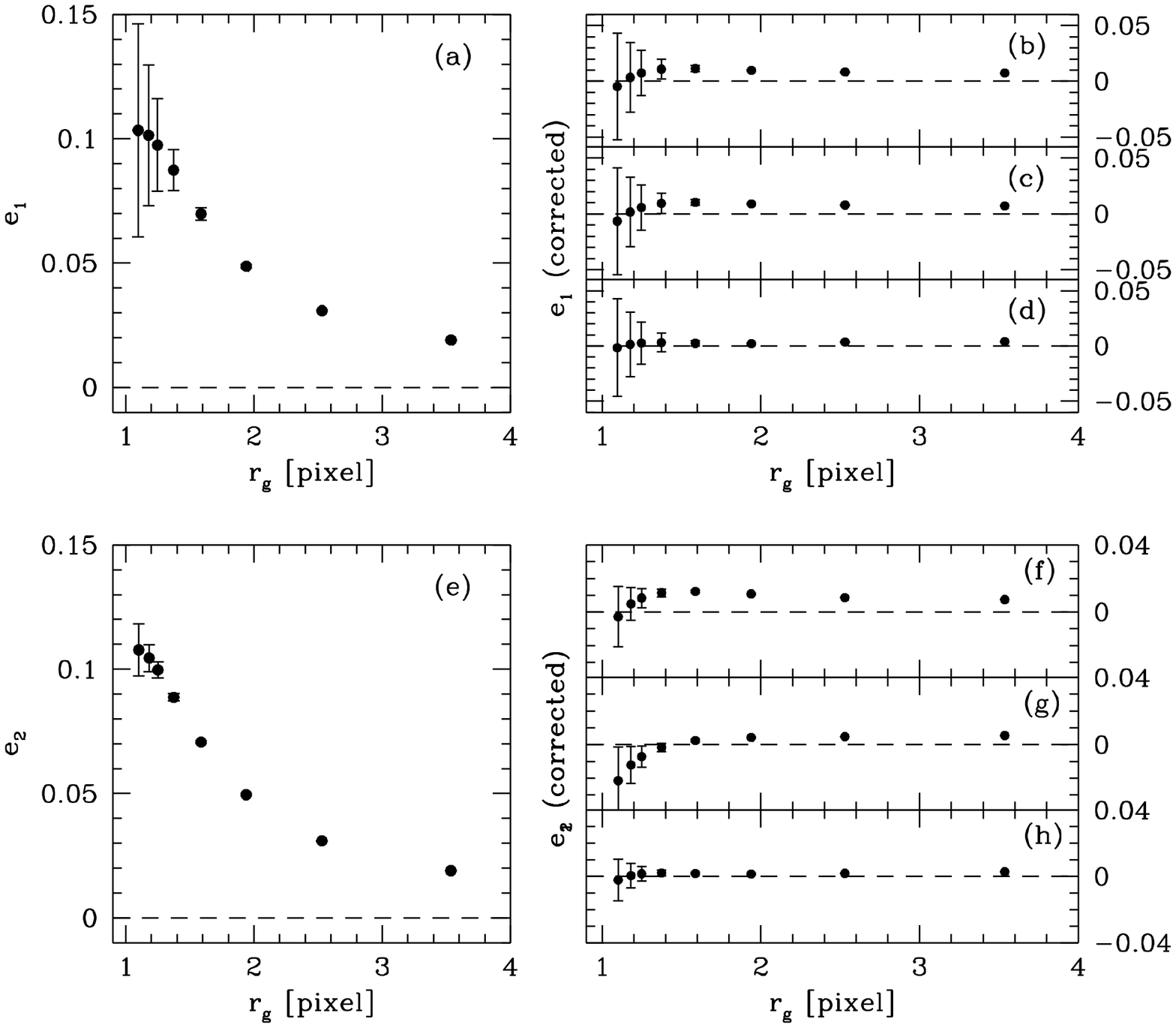}}
\begin{small}
\figcaption{Simulations of the correction for PSF anisotropy. Figure~(a) shows the measured
$e_1$ for objects with exponential profiles, that have an intrinsic $e_1$ of zero as a function of the
size of the object. The errorbars shown indicate the scatter in the measurement
of the polarization due to sampling. Figure~(e) shows the results for
a similar simulation, but now for $e_2$. Figures (b), (c), (d), (f), (g) and (h) show
the corrected polarizations for the various correction schemes. The results shown
in figures~(b) and (f) were obtained using $p_\alpha$ as indicated in KSB95.
Figures~(c) and (g) show the results using unweighted moments for the stars and
(d) and (h) show the results when the weight function for the stars is identical
to that of the corrected object.\label{PSFan}}
\end{small}
\end{center}}

In the simulations we used objects with exponential profiles of different sizes.
These objects were convolved with 10 times oversampled model PSFs, 
calculated by Tiny Tim (Krist \& Hook 1996), block averaged to WPFC2 resolution,
and convolved with the pixel scattering function as given in the Tiny Tim manual.
In order to address the effect of undersampling, objects were placed at different
positions with respect to the pixel grid.

The results are shown in Figure~\ref{PSFan}. The errorbars indicate the scatter
in the polarization measurements due to undersampling effects: clearly
$e_1$ is more sensitive to this problem than $e_2$. 
Figures~\ref{PSFan}(b)--(d) and (f)--(h) show the results of the different PSF anisotropy 
correction schemes. The results in (d) and (h) are obtained using the 
scheme we advocate and used in this paper, and show that this approach performs best.
The simulations show that the undersampling does not cause a bias in the estimation
of the corrected polarization.

\section{'Seeing' correction}

Here we first investigate the correction for PSF induced circularization
presented in LK97. We first consider an almost
'round' PSF, calculated by Tiny Tim (Krist \& Hook 1996). Although 
we apply a small anisotropy correction, this way we can examine the 
LK97 formalism accurately. The anisotropy correction we used 
is the one advocated in this paper.

\vbox{
\begin{center}
\leavevmode
\hbox{%
\epsfxsize=8cm
\epsffile[20 160 580 520]{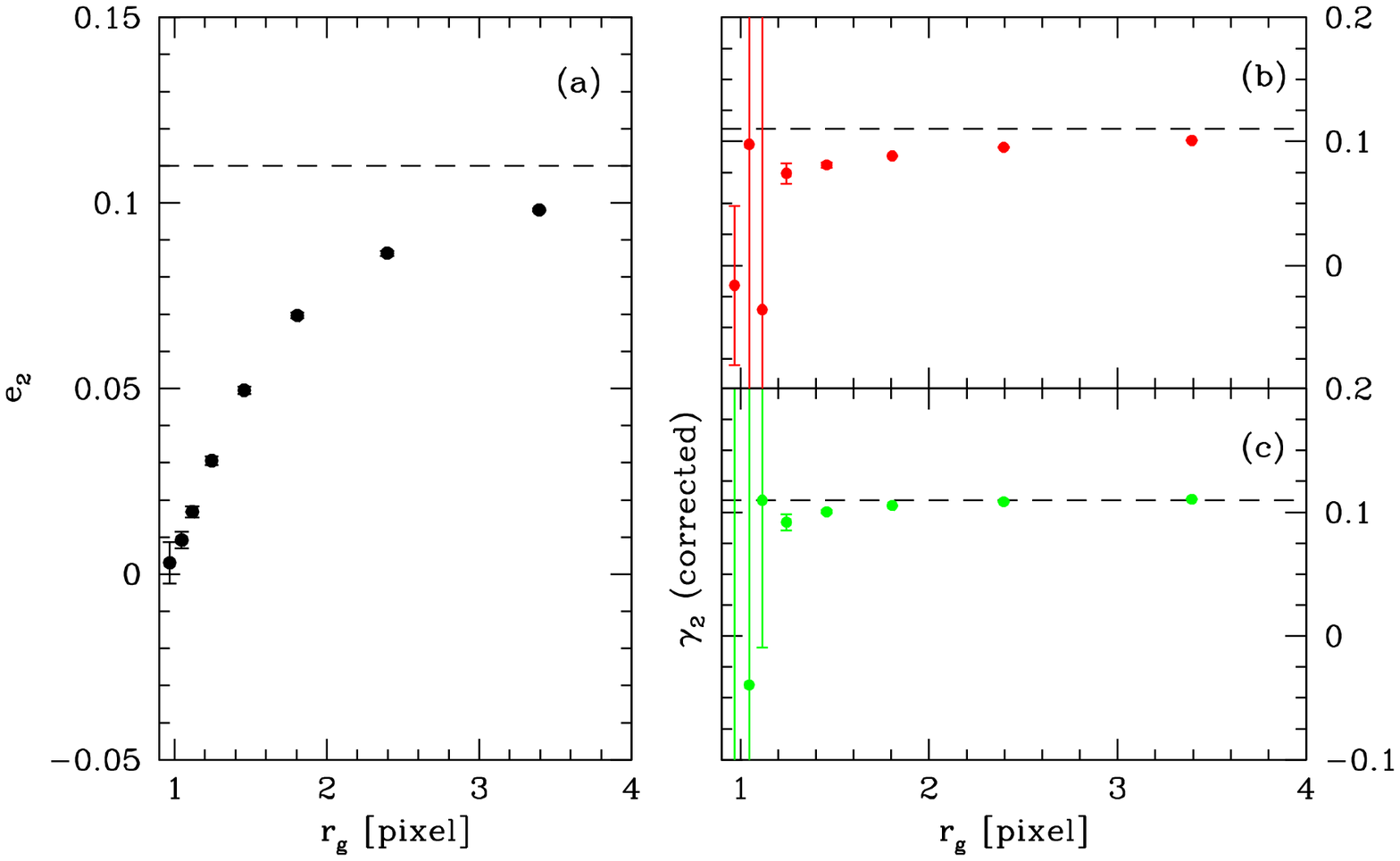}}
\begin{small}
\figcaption{Simulation of the correction for 'seeing', using an almost
circular PSF. Figure~(a) shows the measured $e_2$ for objects with
exponential profiles to which we applied a shear $\gamma_2$ of 11\% (dashed line). 
The errorbars shown indicate the scatter in the measurement
of the polarization due to sampling. Figure~(b) shows the measured shear using
the LK97 formalism. The polarizabilities of the stars were calculated using a
Gaussian weight function with a scale length equal to the scale length of
the stars. Figure~(c) shows the results using a Gaussian weight function with
a scale length equal to the scale length of the corrected object.\label{seeing}}
\end{small}
\end{center}}

If Figure~\ref{seeing} we show the results of such a simulation in which objects
of different sizes are sheared, convolved with the Tiny Tim PSF, and analysed with
the LK97 formula. We applied a
shear $\gamma_2=0.11$. Figure~\ref{seeing}a shows the measured polarizations
as function of the size of the object. Figure~\ref{seeing}b shows the recovered
shear using parameters for the stars, calculated with a constant weight function.
Even for large objects the recovered shear is lower than the applied value. The
results shown in Figure~\ref{seeing}c show that the correction scheme
we adopt in this paper performs best, although the recovered shear is slightly underestimated
for small objects. This might be due to higher order effects in the
calculation of the polarizabilities, which depend on the true shape of the object.
Simulations with shear components $\gamma_1$ show similar results, although the
errorbars for the smallest objects are larger due to undersampling. 

Having shown that both the anisotropy and seeing correction work well independently,
we now show the results of a simulation with an anisotropic PSF, with a position
angle 90 degrees away from the direction of the applied shear. As an additional test,
we added a fair amount of noise to the data.

In Figure~\ref{totalcor} we show the results of this simulation. One sees that
the correction in this case is very large for small objects. Figure~\ref{totalcor}b
shows the results when we calculate the 'pre-seeing' shear polarizability using the 
procedure we advocate, but in the absence of noise. The results in Figure~\ref{totalcor}c 
show that noise does not introduce a significant bias, although the errors for the
smallest objects becomes very large. The shear from the smallest objects is
underestimated slightly, due to the fact that 
the polarizabilities are calculated using an uncorrected value for the polarization.
For small corrections this is a minor effect, but the correction needed in the
simulation shown in Figure~\ref{totalcor} is fairly large. In principle one
could improve the results by correcting the polarizabilities for this effect.

\vbox{
\begin{center}
\leavevmode
\hbox{%
\epsfxsize=8cm
\epsffile[20 160 580 520]{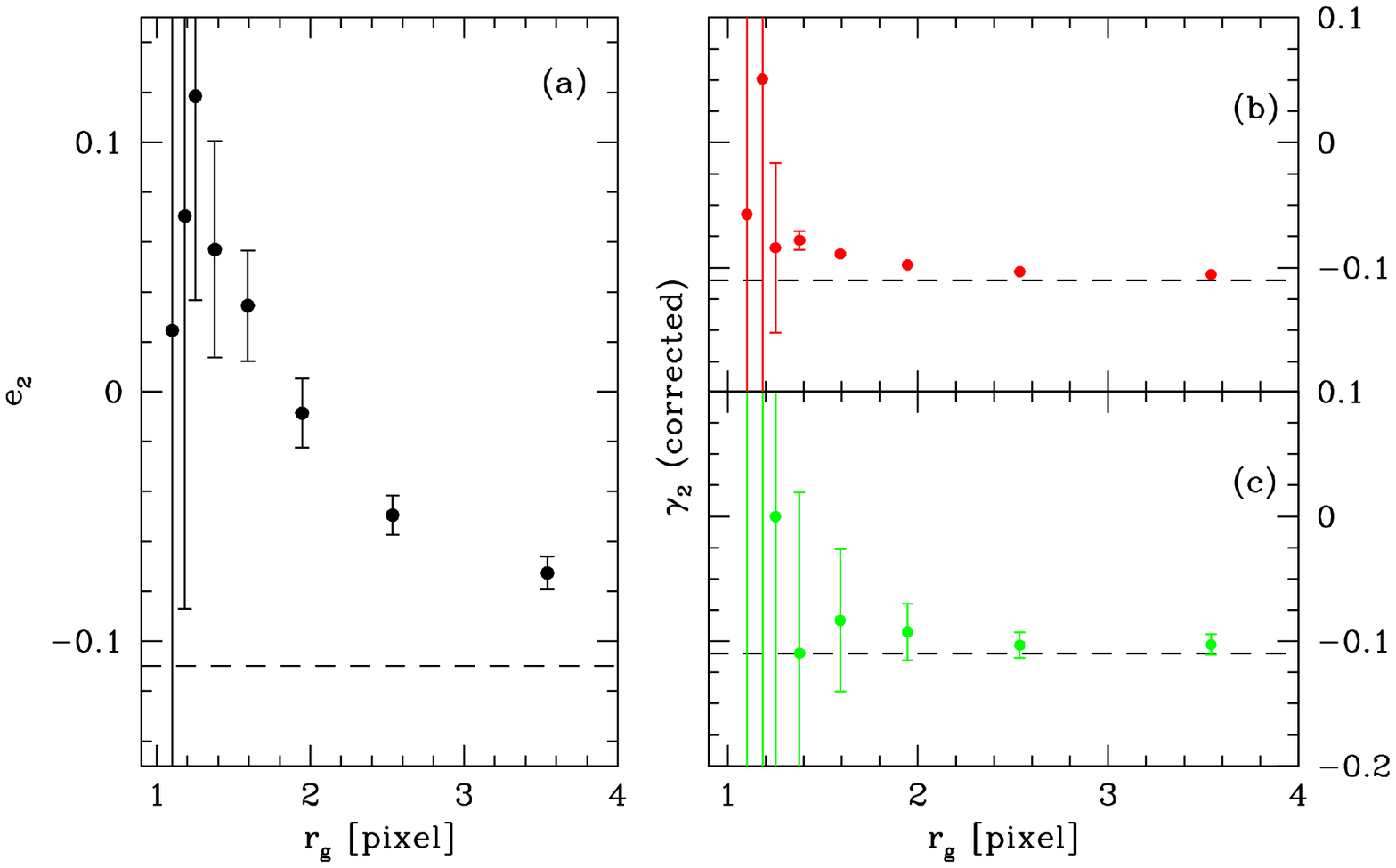}}
\begin{small}
\figcaption{Simulation of the correction for the PSF effects. 
Figure~(a) shows the measured $e_2$ for objects with exponential profiles,
to which we applied a shear $\gamma_2$ of -11\% (dashed line). 
In this simulation, we added noise. The errorbars shown indicate the 
scatter in the measurement of the polarization due to sampling and noise. Figure~(b) 
shows the measured shear in the same simulation, in the absence of noise.
Figure~(c) shows the results in the simulation with noise. In both cases
the results are similar, although the error is very large for small objects
when noise is present.\label{totalcor}}
\end{small}
\end{center}}

Although the shear might be biased low for the smallest objects, we feel that 
we can reliably recover the shear for objects with a scale length $r_g$ larger
than 1.2 pixels.

\section{Analytic results}

As an analytic check on the simulations, we have calculated the
polarizabilities and polarizations for the case when the PSF and the
galaxy image are each the sums of two Gaussians. We have focused on
the (in any case dominant) diagonal terms of the polarizabilities by
considering only shears and polarizations in the $e_1$ component (the
results for $e_2$ follow by symmetry upon rotating the axes by 45
degrees). Thus we only consider Gaussians with principal axes aligned
with the $(x,y)$ coordinates.

As PSF model, we take $[G(s_1,s_1+\delta_1)+G(s_2,s_2+\delta_2)]/2$ where
$G(a,b)$ is a unit integral Gaussian $\exp[-x^2/(2a)-y^2/(2b)]/(2\pi
\sqrt{ab})$ with $x$- and $y$-dispersions $a^{1/2}$ and 
$b^{1/2}$. The pre-seeing, pre-shearing galaxy image is assumed to be
of the form $G(a_1,b_1)+G(a_2,b_2)$.  It is then straightforward to calculate
the axis lengths of this galaxy image after applying a shear $\gamma$:
$a_i\to a_i'=a_i/(1-\gamma)^2$, $b_i\to b_i'=b_i/(1+\gamma)^2$.
Seeing convolution then turns this image into a sum of
four Gaussians:
\begin{equation}
f=\sum_{ij}G(a_i'+s_j,b_i'+s_j+\delta_j)/2.
\end{equation}
While not completely general, these models already allow a large
variety of PSF and galaxy shapes to be simulated: in particular,
radially varying ellipticity profiles are possible.

\vbox{
\begin{center}
\leavevmode
\hbox{%
\epsfxsize=8cm
\epsffile{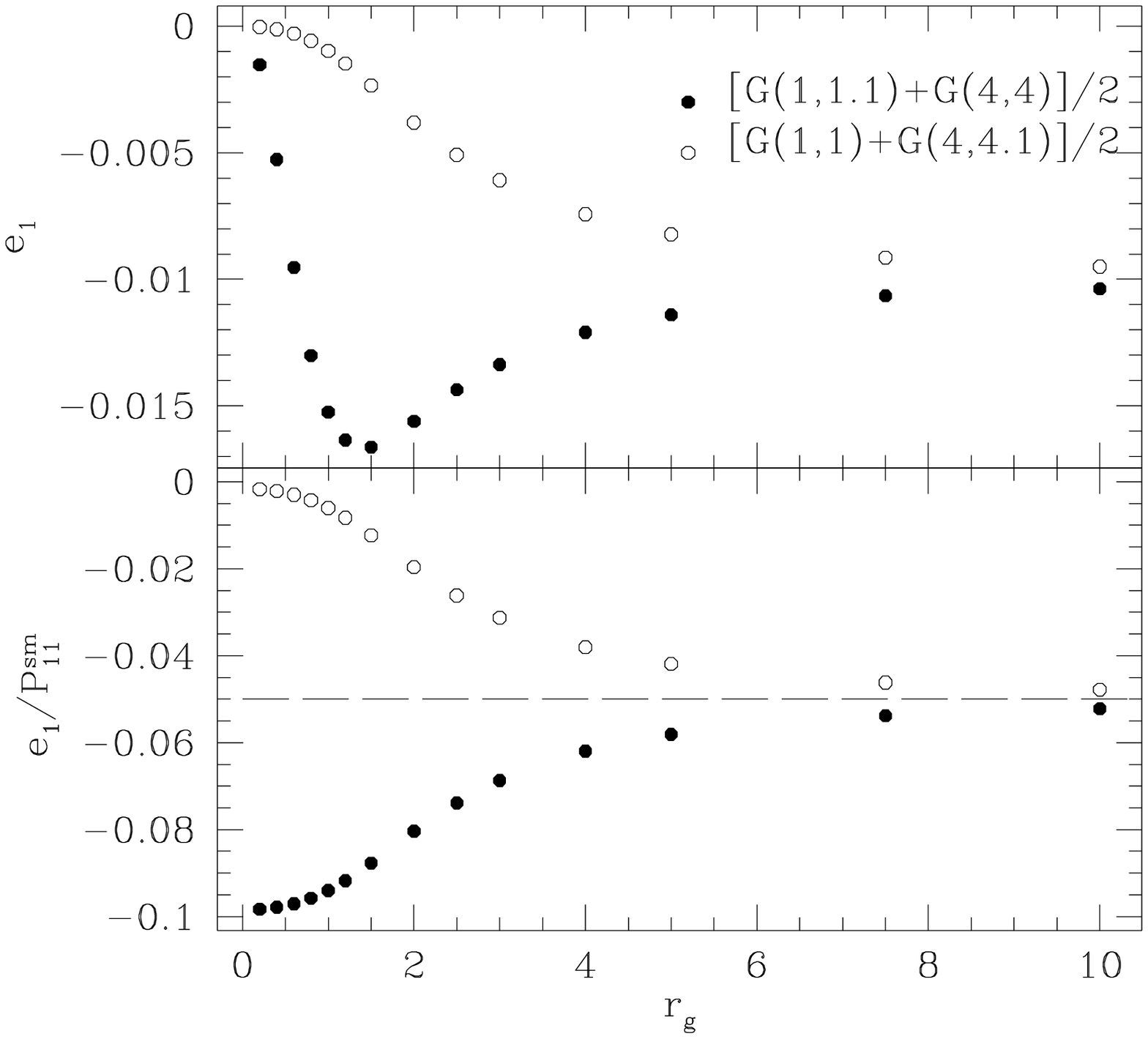}}
\begin{small}
\figcaption{The measured polarization $e_1$ (top panel), obtained with a
Gaussian weight function of dispersion $r_g$, and the deduced
estimate of the intrinsic polarization $p_1$ (bottom) for two
double-Gaussian stellar profiles with identical second monents (and
hence identical $p_1$). It is clear that the choice of weight function
is important in the correction for PSF anisotropy.\label{fig:starpolar}}
\end{small}
\end{center}}

We have used symbolic mathematics to compute the polarizabilities and
polarizations of the PSF and of the final observed image, assuming
that the PSF anisotropy parameters $\delta_1$ and $\delta_2$ and the
shear $\gamma$ are small. A circular Gaussian weight function, of
dispersion $r_g$, was used in these calculations.

First we consider the measurement of the PSF anisotropy. In
Figure~\ref{fig:starpolar} we compare the estimate $e_1/P^{\rm
sm}_{11}$ of the PSF polarization for two PSF's with identical smear
polarizability (to leading order in $\delta_i$) and second moments,
but different ellipticity profiles.  The plot shows that the choice of
weight function is important: since the measured $e_1$ is different
for the two models, it is impossible for the smear polarizability to
correct both cases accurately. As may be seen, $|p_1|$ is
systematically overestimated for a PSF with radially decreasing
ellipticity.

What is the impact of these results on PSF anisotropy correction of
galaxy polarizations? To answer this question, we have simulated the
KSB95 procedure on an anisotropic double-Gaussian galaxy (see
Figure~\ref{fig:PSFcor}). We picked galaxy models with varying
ellipticity profiles, convolved with the two PSF's considered above, and
calculated the resulting polarization shifts, $\delta e_1$ and smear
polarizabilities $P_{11}^{\rm sm}({\rm gal})$ using a Gaussian weight
function whose size matched the size of the galaxy image. This
polarization shift was then corrected for PSF anisotropy following the
KSB method, where when measuring the stellar images we allowed
$r_g(\rm star)$ to vary.  

Evidently the correction depends on the
stellar images' weight function, and a good compromise appears to be
to take the stellar and the galaxy weight functions the
same. Intuitively this result is perhaps not surprising: it seems
sensible to weigh the radial ellipticity profiles of galaxy and PSF in
a similar fashion.

We have also examined the LK97 formalism for correcting for
seeing convolution as a function of the size of the weight functions.
Figure~\ref{fig:lksim} shows the LK estimator $\delta
e_1/P^\gamma_{11}$ for the shear, where again in $P^\gamma=P^{\rm
sh}{\rm(gal)}-P^{\rm sh}{\rm(star)}P^{\rm sm}{\rm(gal)}/P^{\rm
sm}{\rm(star)}$ the stellar polarizabilities have been computed over a
range of different $r_g$. Here large dependences on the stellar weight
function are evident, and we conclude that the best strategy is to use
the same Gaussian weight function for the stellar and for the galaxy
images.

\vbox{
\begin{center}
\leavevmode
\hbox{%
\epsfxsize=8cm
\epsffile{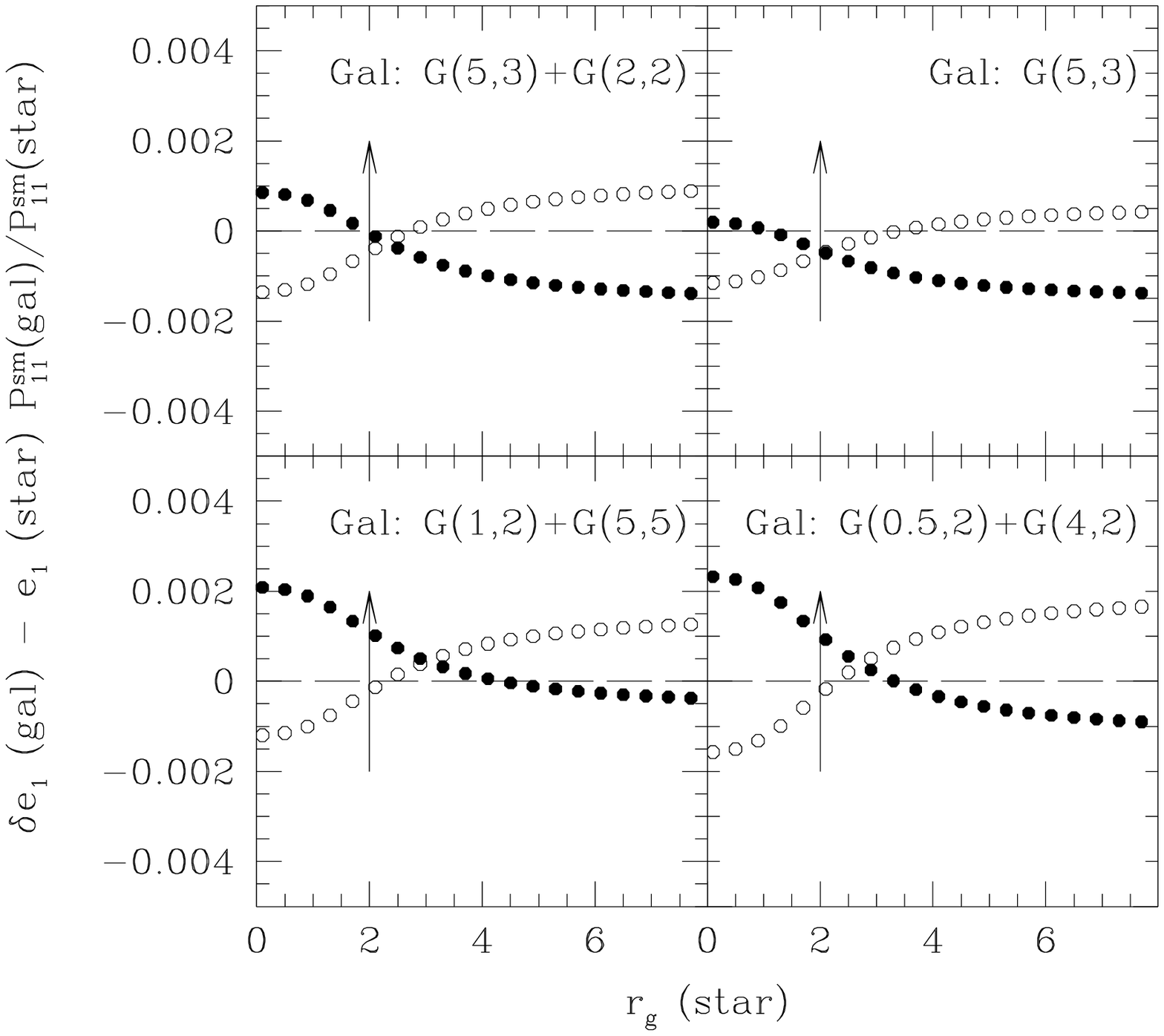}}
\begin{small}
\figcaption{Simulations of the PSF anisotropy correction for 
double-Gaussian galaxies smeared with the two PSF's considered in
Figure~\ref{fig:starpolar}. The galaxy image properties were measured
with a Gaussian weight function of dispersion $r_g=2$ (indicated by
the vertical arrow), and the stellar profiles with $r_g$ ranging from
0.1 to 8. The variances of the Gaussians making up the galaxy image
are indicated in each panel: $G(a,b)$ stands for a unit-integral
Gaussian of $\overline{x^2}=a$, $\overline{y^2}=b$. Circularly-smeared
galaxy polarizations are (clockwise from top left) 0.033, 0.068,
$-$0.042, $-$0.022. In each case, choosing a stellar $r_g$ smaller
than that used for measuring the galaxy image performs worse than when
the stellar and galaxy $r_g$ are similar.\label{fig:PSFcor}}
\end{small}
\end{center}}

\vbox{
\begin{center}
\leavevmode
\hbox{%
\epsfxsize=8cm
\epsffile{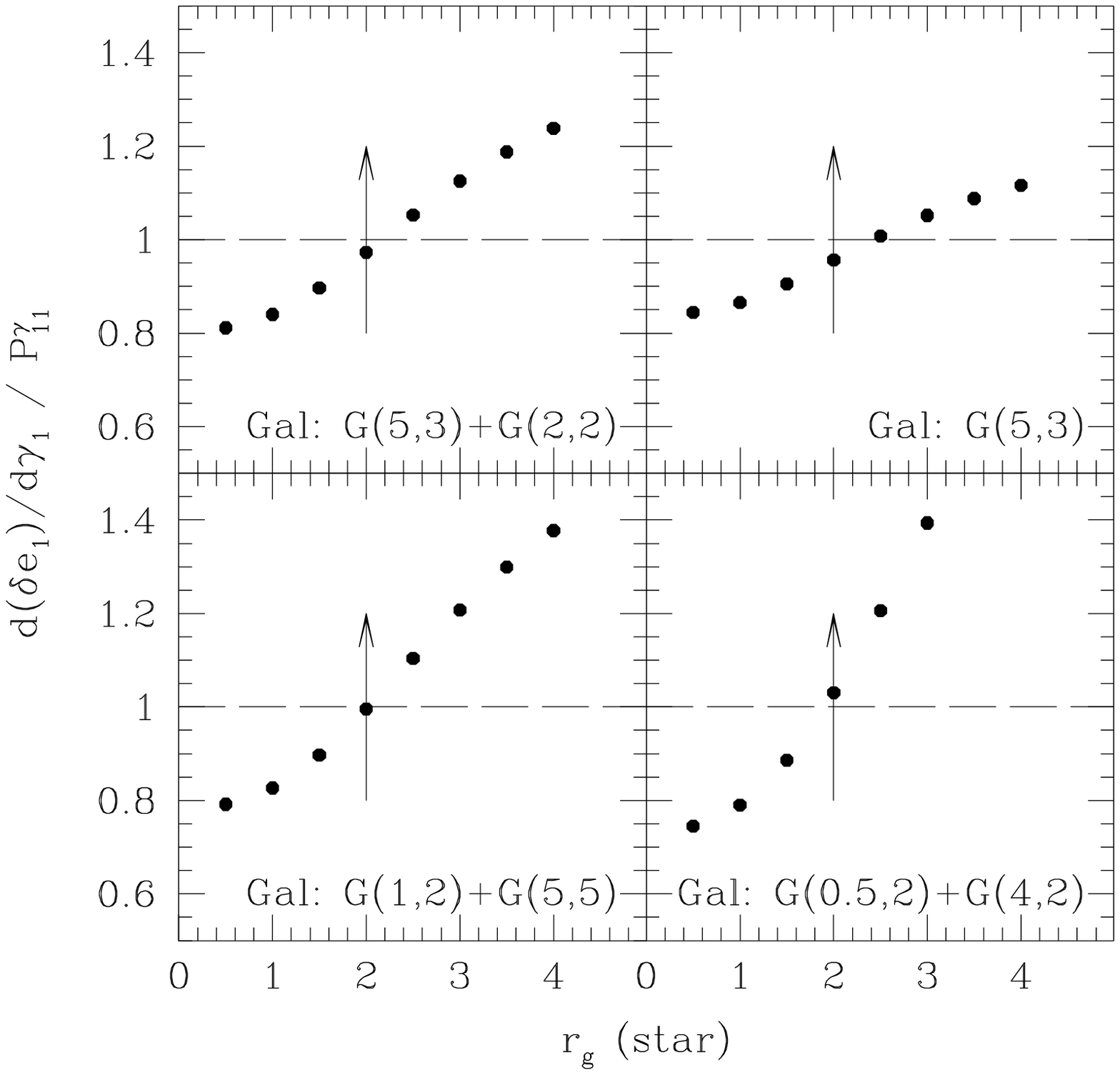}}
\begin{small}
\figcaption{Simulations of the shear polarizability correction using the
Luppino-Kaiser formalism. The given galaxy models are sheared by a
(small) shear $\gamma_1$, and then smeared with the circular
double-Gaussian PSF $[G(1,1)+G(2,2)]/2$. Then the Luppino-Kaiser
pre-seeing shear polarizability $P^\gamma_{11}$ is calculated using a
Gaussian weight function of dispersion 2 (indicated by the arrows) for
the galaxy images, and a range of dispersions $r_g$ for the stellar
PSF. The plot shows the ratio between the actual pre-seeing
polarizability $d\delta e_1/d\gamma_1$ and the LK $P_{11}^\gamma$.\label{fig:lksim}}
\end{small}
\end{center}}
	
\end{document}